\documentclass[twocolumn]{aastex7}
\shorttitle{GW241011 and GW241110}
\shortauthors{Lei He et al.}
\graphicspath{{./}}
\usepackage{times}
\usepackage{amsmath}
\begin{document}
\title{Exploring Hierarchical Merger Scenarios for GW241011 and GW241110}
\correspondingauthor{Lei He, Wen Zhao}
\author[0000-0001-7613-5815]{Lei He}
\affiliation{Department of Astronomy, University of Science and Technology of China, Hefei, Anhui 230026, People’s Republic of China}
\affiliation{School of Astronomy and Space Sciences, University of Science and Technology of China, Hefei, Anhui 230026, People’s Republic of China}
\email[show]{helei0831@mail.ustc.edu.cn}

\author[0000-0001-7688-6504]{Liang-Gui Zhu}
\affiliation{Department of Astronomy, University of Science and Technology of China, Hefei, Anhui 230026, People’s Republic of China}
\affiliation{School of Astronomy and Space Sciences, University of Science and Technology of China, Hefei, Anhui 230026, People’s Republic of China}
\email{lianggui.zhu@ustc.edu.cn}

\author[0000-0002-2242-1514]{Zheng-Yan Liu}
\affiliation{Department of Astronomy, University of Science and Technology of China, Hefei, Anhui 230026, People’s Republic of China}
\affiliation{School of Astronomy and Space Sciences, University of Science and Technology of China, Hefei, Anhui 230026, People’s Republic of China}
\email{ustclzy@mail.ustc.edu.cn}

\author[0000-0001-9098-6800]{Rui Niu}
\affiliation{Department of Astronomy, University of Science and Technology of China, Hefei, Anhui 230026, People’s Republic of China}
\affiliation{School of Astronomy and Space Sciences, University of Science and Technology of China, Hefei, Anhui 230026, People’s Republic of China}
\email{nrui@ustc.edu.cn}

\author[0009-0004-8038-4741]{Chao Wei}
\affiliation{Department of Astronomy, University of Science and Technology of China, Hefei, Anhui 230026, People’s Republic of China}
\affiliation{School of Astronomy and Space Sciences, University of Science and Technology of China, Hefei, Anhui 230026, People’s Republic of China}
\email{charles272@mail.ustc.edu.cn}

\author[0000-0001-8955-0452]{Ken Chen}
\affiliation{Department of Astronomy, University of Science and Technology of China, Hefei, Anhui 230026, People’s Republic of China}
\affiliation{School of Astronomy and Space Sciences, University of Science and Technology of China, Hefei, Anhui 230026, People’s Republic of China}
\email{cken@ustc.edu.cn}

\author[0000-0002-1330-2329]{Wen Zhao}
\affiliation{Department of Astronomy, University of Science and Technology of China, Hefei, Anhui 230026, People’s Republic of China}
\affiliation{School of Astronomy and Space Sciences, University of Science and Technology of China, Hefei, Anhui 230026, People’s Republic of China}
\affiliation{College of Physics, Guizhou University, Guiyang, 550025, People’s Republic of China}
\email[show]{wzhao7@ustc.edu.cn}

\begin{abstract}
  GW241011 and GW241110 are asymmetric binary black hole mergers with rapidly spinning primaries, unequal component masses, and nonzero spin--orbit tilts, making them natural candidates for hierarchical mergers. We use a Bayesian framework to compare a fiducial first-generation (1G+1G) binary black hole population with second-generation plus first-generation (2G+1G) hierarchical merger models in star clusters and active galactic nucleus (AGN) disks. Both events favor the 2G+1G interpretation over the 1G+1G hypothesis, with $\ln\mathcal{B}^{\rm 2G+1G}_{\rm 1G+1G}\simeq6.5$--$8.6$ for GW241011 and $\ln\mathcal{B}^{\rm 2G+1G}_{\rm 1G+1G}\simeq3.0$--$4.5$ for GW241110, depending on the waveform model and assumed environment. The AGN disk models yields slightly larger evidence than the star cluster models, mainly due to their spin tilt distribution, but the data do not provide a decisive environmental classification. We further consider a third-generation plus first-generation (3G+1G) interpretation, but it is not robustly preferred over 2G+1G scenarios. Finally, we also search for optical counterparts by examining AGNs within the three-dimensional localization volumes using ZTF and ATLAS forced photometry, and find one candidate source with weak flare, which might be associated with GW241110 event. 
\end{abstract}


\section{Introduction}
The recently released Gravitational-Wave Transient Catalog version 5.0 (GWTC-5.0) by LIGO--Virgo--KAGRA (LVK) Collaboration has substantially expanded the observed population of binary black hole (BBH) mergers, incorporating events from the second part of the fourth observing run (O4b) \citep{collaborationGWTC50ObservationsSecond2026,collaborationGWTC50IntroductionVersion2026}. Among the BBH events in GWTC-5.0, GW241011 and GW241110 stand out as two asymmetric, high-spin systems and have been highlighted in a dedicated study \citep{abacGW241011GW241110Exploring2025}. Both events contain a rapidly spinning primary black hole (BH), unequal component masses, and nonzero primary spin--orbit tilts. For GW241011, the inferred primary dimensionless spin magnitude is $\chi_1=0.78^{+0.09}_{-0.09}$, with a spin--orbit tilt angle of $\theta_1=31^{+11}_{-14} \deg$ and a mass ratio of $q=0.30^{+0.09}_{-0.08}$. For GW241110, the corresponding inferred values are $\chi_1=0.61^{+0.33}_{-0.40}$, $\theta_1=133^{+47}_{-25} \deg$, and $q=0.45^{+0.32}_{-0.17}$ \citep[see Table~1 in ][]{abacGW241011GW241110Exploring2025}.

The combination of large primary spin magnitudes, significant spin--orbit tilt angles and unequal component masses is difficult to accommodate within the standard isolated binary evolution channel for the first-generation (1G) BBHs, which generally predicts relatively small natal BH spins and spin orientations preferentially aligned with the binary orbital angular momentum \citep{zaldarriagaExpectedSpinsGravitational2018,qinSpinSecondbornBlack2018,fullerMostBlackHoles2019,baveraImpactMasstransferPhysics2021}. These features motivate the consideration of alternative formation pathways, and hierarchical mergers provide one natural alternative \citep{mapelliFormationChannelsSingle2021,abacGW241011GW241110Exploring2025}. In this scenario, at least one component BH is the remnant of a previous BBH merger. If the merger remnant is retained in its host environment, it can subsequently pair with another BH and merge again, producing a higher-generation BH. Such merger remnants are expected to be more massive than typical 1G BHs and to have relatively large dimensionless spins around 0.7 \citep{pretoriusEvolutionBinaryBlackHole2005,gerosaHierarchicalMergersStellarmass2021}. A merger between a higher-generation BH and a lower-generation companion can therefore naturally produce a system with a rapidly spinning primary and an unequal mass ratio, similar to the properties inferred for GW241011 and GW241110. Recent population work has identified GW241011 and GW241110 as candidate members of a low-mass, rapidly spinning subpopulation that is consistent with dynamical assembly and a hierarchical merger origin \citep{tongSubpopulationLowmassSpinning2025}. An event-focused analysis likewise found GW241011 to be consistent with a hierarchical merger origin in dense star clusters, while noting that the interpretation of GW241110 remains under debate due to its broader parameter uncertainties \citep{liGW241011GW241110Hints2025}.

Hierarchical mergers are expected to occur in dense environments, such as star clusters (e.g., nuclear clusters and globular clusters) and active galactic nuclei (AGN) disks \citep{gerosaHierarchicalMergersStellarmass2021,mandelRatesCompactObject2022,liComparingHierarchicalBlack2023,liOriginChannelsHierarchical2025,zhuEvidenceFractionLIGO2025a}. In star clusters, BBHs can be assembled dynamically through binary--single encounters or three-body processes \citep{heggieBinaryEvolutionStellar1975}. A merger remnant can participate in a subsequent merger only if its gravitational-wave recoil velocity is smaller than the escape velocity of the host system. Since dynamically assembled binaries do not have a preferred orientation, their spin directions are generally expected to be approximately isotropic \citep{liComparingHierarchicalBlack2023}. In contrast, AGN disks provide a gas-rich environment in which BHs can be captured into the disk, migrate, form binaries and merge efficiently \citep{bellovaryMIGRATIONTRAPSDISKS2016,bartosRapidBrightStellarmass2017,tagawaFormationEvolutionCompactobject2020,vaccaroImpactGasHardening2024a,vaccaroAGNdrivenBBHMergers2026}. The existence of the central supermassive black hole (SMBH) can help retain merger remnants within the AGN environment, potentially leading to a larger fraction of hierarchical mergers in AGN disks than in star clusters \citep{liComparingHierarchicalBlack2023}. In addition, because the gaseous disk defines a preferred orbital plane, the spins of higher-generation BHs inherited from the previous mergers may preferentially appear aligned or anti-aligned with the orbital angular momentum of subsequent BBH mergers \citep{bartosRapidBrightStellarmass2017,yangHierarchicalBlackHole2019,mckernanMonteCarloSimulations2020}. 

If a BBH merger occurs in an AGN disk, the gaseous environment may enable observable electromagnetic (EM) emission \citep{bartosRapidBrightStellarmass2017}, unlike in other gas-poor formation channels. Such emission is generally expected to arise from the interactions between the binary or its merger remnant and the surrounding gas. Proposed mechanisms include accretion-powered outflows \citep{wangAccretionmodifiedStarsAccretion2021,kimuraOutflowBubblesCompact2021,rodriguez-ramirezOpticalUVFlares2025}, jet-driven emission \citep{tagawaShockCoolingBreakout2024,rodriguez-ramirezOpticalEmissionModel2023a,chenElectromagneticCounterpartsPowered2024,chenObservationalPropertiesThermal2025,chenObservationalPropertiesNonthermal2026}, and ram pressure stripping of the surrounding gas \citep{mckernanRampressureStrippingKicked2019}. The identification of such an EM signal would provide strong evidence for an AGN disk origin. However, establishing a physical association remains challenging because AGNs are intrinsically variable and the GW localization regions are typically large. Several candidates have been reported by the three-dimensional alignment between AGN and GW events
\citep{grahamCandidateElectromagneticCounterpart2020,grahamLightDarkSearching2023,cabreraSearchingElectromagneticEmission2024b,heSearchingElectromagneticCounterpart2025,zhuConstrainingFractionLIGO2026}. However, future multiband observations are still needed for definitive confirmation \citep{ashtonCurrentObservationsAre2021,palmeseLIGOVirgoBlack2021a,mortonGW190521BinaryBlack2023,heTracingLightIdentification2025}, which are also crucial to clarify the physical interpretation of the EM emissions.

In this work, we adopt a Bayesian framework similar to that of \citet{liHierarchicalMergerScenario2025} to investigate the hierarchical merger scenarios for GW241011 and GW241110. In Section~\ref{sec:method}, we describe the methodology used to assess their formation channels and the construction of the first- and higher-generation population models. We present the results in Section~\ref{sec:results}. In Section~\ref{sec:search}, we describe our search for EM counterparts associated with these two events. Finally, we discuss the implications of our results and draw conclusions in Section~\ref{sec:conclusion}. Throughout this work, we adopt the cosmological parameters reported by \citet{adePlanck2015Results2016}.

\section{Method\label{sec:method}}
\subsection{Bayesian Framework}

To quantify the relative support for different formation scenarios of a GW event, we compute the Bayes factor between competing hypotheses, defined as

\begin{equation}
  \mathcal{B}^i_j(d) = \frac{p(d|\mathcal{H}_i)}{p(d|\mathcal{H}_j)},
\end{equation}
where $d$ denotes the GW data, and $\mathcal{H}_i$ and $\mathcal{H}_j$ represent two competing hypotheses. The Bayes factor therefore measures how strongly the observed data favor one scenario over another. 

In this work, we consider several binary black hole formation scenarios, including first-generation mergers (1G+1G), and hierarchical mergers in dense star clusters or AGN disks.

\subsection{Evaluation of Evidence}
For each scenario $\mathcal{H}_i$, the Bayesian evidence is computed by marginalizing the likelihood over the model parameter space,
\begin{equation}
  p(d|\mathcal{H}_i) = \int p(d|\boldsymbol{\lambda}) \, p(\boldsymbol{\lambda} | \mathcal{H}_i) \mathrm{d}\boldsymbol{\lambda},
  \label{eq:evidence}
\end{equation}
where $\boldsymbol{\lambda}$ denotes the set of parameters, $p(d|\boldsymbol{\lambda})$ is the likelihood of the observed GW data, and $p(\boldsymbol{\lambda} | \mathcal{H}_i)$ represents the prior distribution predicted by the corresponding formation scenario. The evidence quantifies the overall consistency between the observed data and a given formation scenario. In this work, the parameter set $\boldsymbol{\lambda}$ includes the source-frame primary mass $m_1$, the dimensionless spin magnitude of the primary BH $\chi_1$, the primary spin tilt angle cosine $\cos\theta_1$ between the primary spin and the orbital angular momentum, and mass ratio $q$.

The likelihood can be reconstructed from the posterior samples through
\begin{equation}
p(d | \boldsymbol{\lambda}) \propto \frac{p(m_1, q, \chi_1,\cos\theta_1 | d)}{p(m_1, q, \chi_1,\cos\theta_1)},  
\end{equation}
where $p(m_1, q, \chi_1,\cos\theta_1 | d)$ is the posterior distribution obtained from GW parameter estimation, and $p(m_1, q, \chi_1,\cos\theta_1)$ is the prior distribution adopted in the original parameter estimation analysis, which is uniform in detector-frame component masses and spin magnitudes, and isotropic in spin orientations \citep{abacGW241011GW241110Exploring2025}.
  
The posterior distributions are reconstructed from the GW posterior samples released by the LVK Collaboration \footnote{\url{https://zenodo.org/records/17343574}} using the \texttt{FIGARO} package \citep{rinaldiFIGAROHierarchicalNonparametric2024}, which employs a Dirichlet process Gaussian mixture model \citep[DPGMM,][]{nguyenApproximationFiniteMixtures2020} to flexibly approximate multidimensional probability densities. 

Similarly, for each scenario, the prior distribution is constructed as 
\begin{equation}
  p(\boldsymbol{\lambda} | \mathcal{H}_i) = p(m_1, q, \chi_1, \cos\theta_1 | \mathcal{H}_i),
\end{equation}
which describes the distribution of these parameters. The detailed construction of the prior distributions is described in the following subsection. The evidence integral in Equation~\ref{eq:evidence} is evaluated with \texttt{dynesty} \citep{speagleDynestyDynamicNested2020}, as implemented in \texttt{bilby} \citep{ashtonBilbyUserfriendlyBayesian2019}.

\subsection{Population Priors for Formation Hypotheses}
\label{sec:population_prior}
  
In this subsection, we describe how the population priors are constructed for the formation hypotheses considered in this work. These priors encode the main astrophysical differences between different formation channels.

\subsubsection{First-generation mergers}

For the first-generation binary hypothesis ($\mathcal{H}_{\rm 1G+1G}$), we adopt the population distributions inferred by the LVK Collaboration from GWTC-4.0 under the \textsc{Broken Power Law + 2 Peaks} mass model and \textsc{Gaussian Component Spins} model \citep{collaborationGWTC40PopulationProperties2025}, similar to \citet{liOriginChannelsHierarchical2025}. Although these distributions characterize the overall BBH population and may include contributions from mergers involving higher-generation BHs, such mergers are expected to constitute only a small fraction of the observed BBH population \citep{liResolvingStellarCollapseHierarchicalMerger2024,liRevealingHeffCorrelation2025,liAlignedHierarchicalBlack2026,liOriginChannelsHierarchical2025,gayathriBlackHoleMergers2021}, making this an adequate approximation for the 1G+1G population.

The LVK population analysis provides posterior samples of the one-dimensional marginal population distributions for the primary mass, primary spin magnitude, primary spin orientation, and mass ratio. We use the median posterior predictive distribution for each parameter as our fiducial 1G+1G population prior. Motivated by the expected pair-instability mass gap \citep{woosleyPulsationalPairinstabilitySupernovae2017,tongEvidencePairinstabilityGap2026}, we truncate the primary mass prior at $m_1=65\, M_\odot$. The prior for the first-generation binaries is therefore written as 
\begin{equation}
  \begin{aligned}
  p(m_1, q,& \chi_1, \cos\theta_1 | \mathcal{H}_{\rm 1G+1G}) \\
   \approx &\, p_{\rm LVK}(m_1)\, p_{\rm LVK}(q)\, p_{\rm LVK}(\chi_1)\, p_{\rm LVK}(\cos\theta_1),  
  \end{aligned}
\end{equation}
where $p_{\rm LVK}$ denotes the corresponding prior distribution inferred from the LVK analysis.

\subsubsection{Hierarchical mergers}

For the hierarchical-merger hypotheses, at least one component black hole is assumed to be the remnant of one or more previous BBH mergers. We construct the corresponding population priors recursively, starting from the fiducial 1G+1G population described above.

We first generate $N_{1G}=10^6$ 1G+1G binaries from the fiducial 1G+1G prior. For each binary, the sampled component masses, spin magnitudes, and spin orientations are used to compute the properties of the merger remnant, including its mass, spin, and recoil velocity. This calculation is performed using the numerical-relativity remnant surrogate model \texttt{NRSur7dq4Remnant} \citep{varmaSurrogateModelsPrecessing2019}, which provides remnant predictions for quasi-circular BBH mergers. The merger remnants obtained in this step are treated as candidate second-generation (2G) black holes.

Whether a candidate 2G BH can participate in a subsequent merger depends on whether it is retained in its host environment after receiving the recoil kick. For the star cluster channel, we retain a remnant only if its recoil velocity is smaller than the host escape velocity. The escape velocity is typically of order a few tens of $\mathrm{km\,s^{-1}}$ for globular clusters and a few hundred $\mathrm{km\,s^{-1}}$ for nuclear clusters \citep{antoniniMERGINGBLACKHOLE2016}. We adopt a fiducial value $v_{\rm esc} = 100\, \mathrm{km\, s^{-1}}$ in this work. For the AGN disk channel, we assume that all merger remnants are retained and can participate in subsequent mergers.

We then construct the 2G+1G population by pairing a retained 2G remnant with a companion drawn from the fiducial 1G population. At this stage, the component masses and spin magnitudes are specified, as is the spin-orbit tilt of the 1G BH. The spin-orbit tilt of the 2G remnant is assigned according to the formation environment. 

For the star cluster channel, we assume that the spin orientation of the 2G BH is isotropic with respect to the orbital angular momentum of the newly formed binary. For the AGN disk channel, we follow the prescription of \citet{yangHierarchicalBlackHole2019} and assume that the orbital axis of each BBH is aligned with the AGN disk. Because the spins of 1G BHs are generally small, the spin of a 2G remnant is expected to be dominated by the orbital angular momentum of its progenitor binary and thus to be approximately aligned with the AGN disk axis. We further assume that this spin direction is preserved until the subsequent merger. Therefore, the spin of the 2G BH is expected to be either aligned or anti-aligned with the orbital angular momentum of the newly formed binary.

\begin{figure*}[htb!]

  \gridline{
    \includegraphics[width=0.48\textwidth]{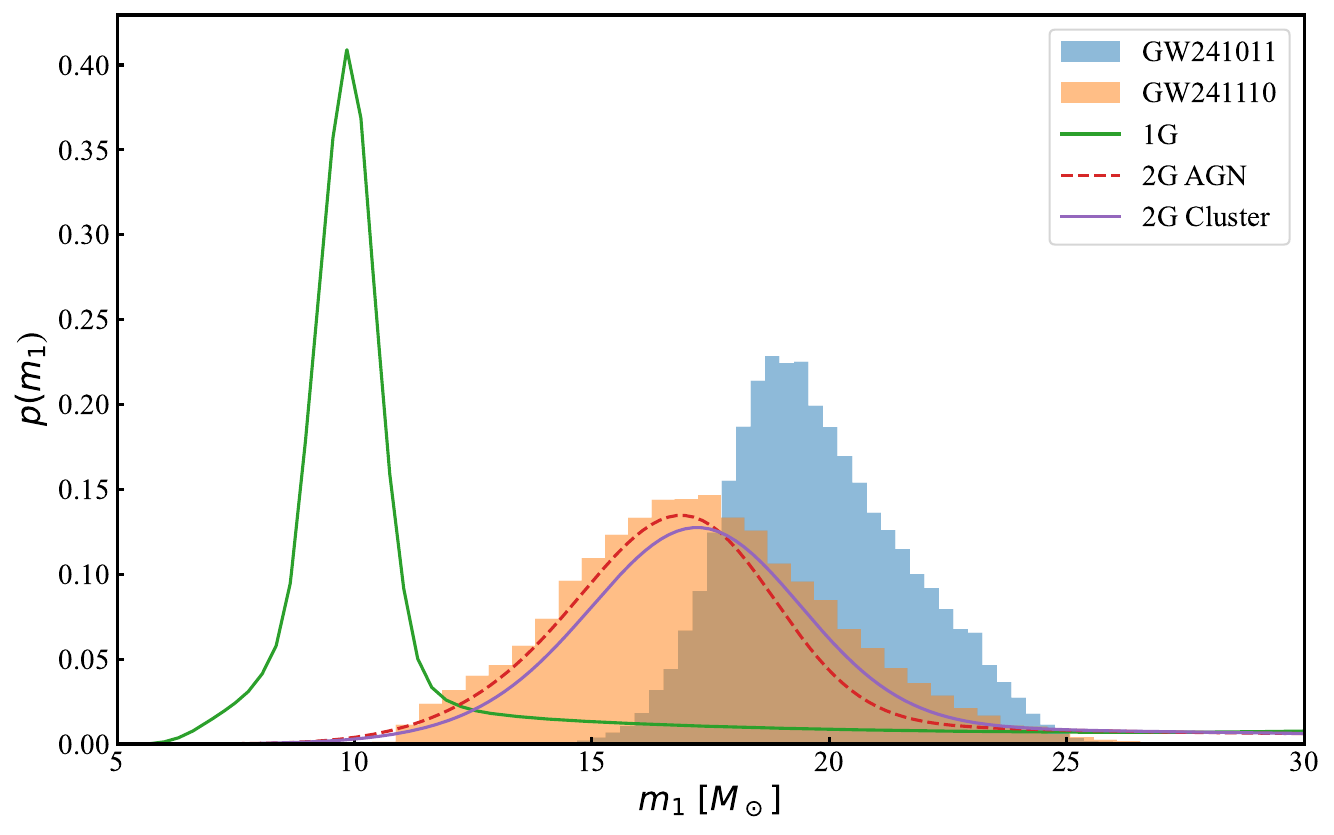}
    \includegraphics[width=0.48\textwidth]{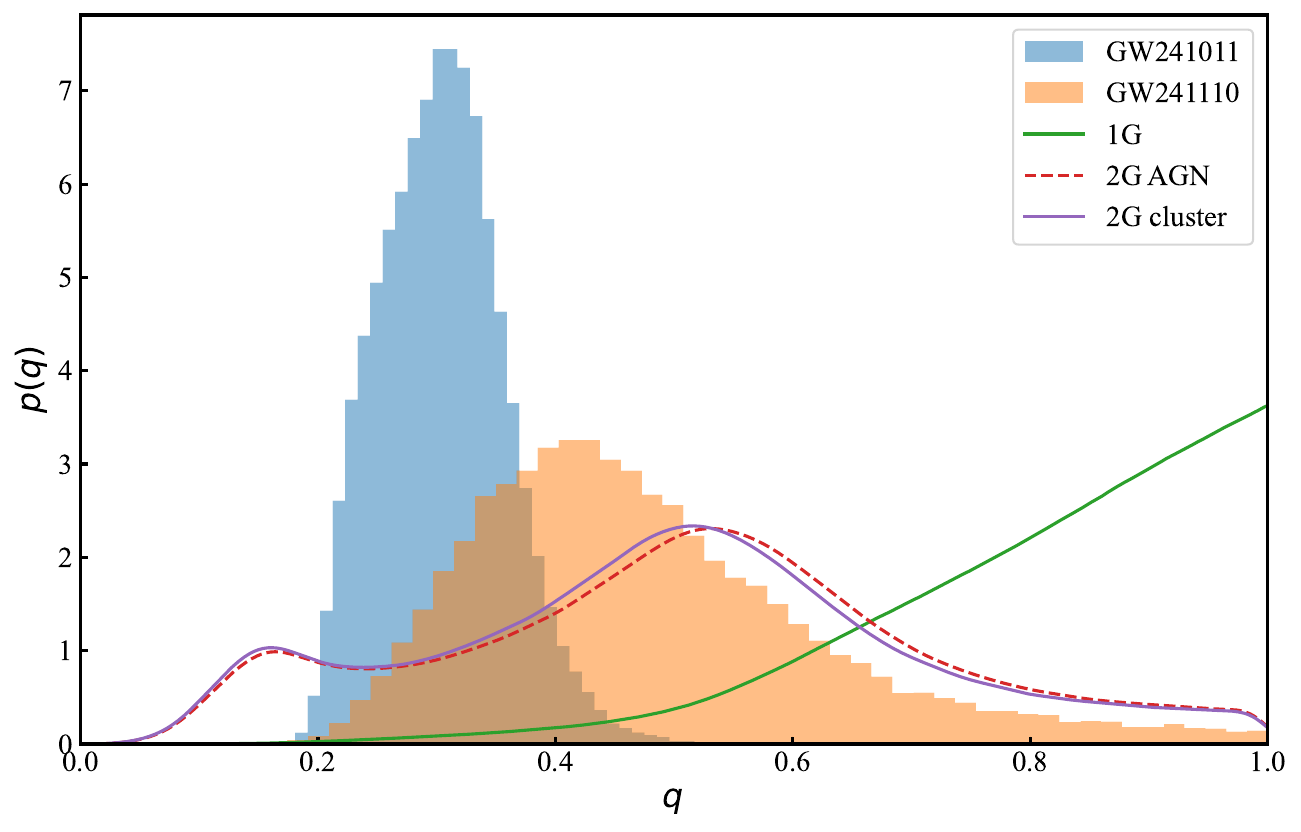}
  }
  \gridline{
    \includegraphics[width=0.48\textwidth]{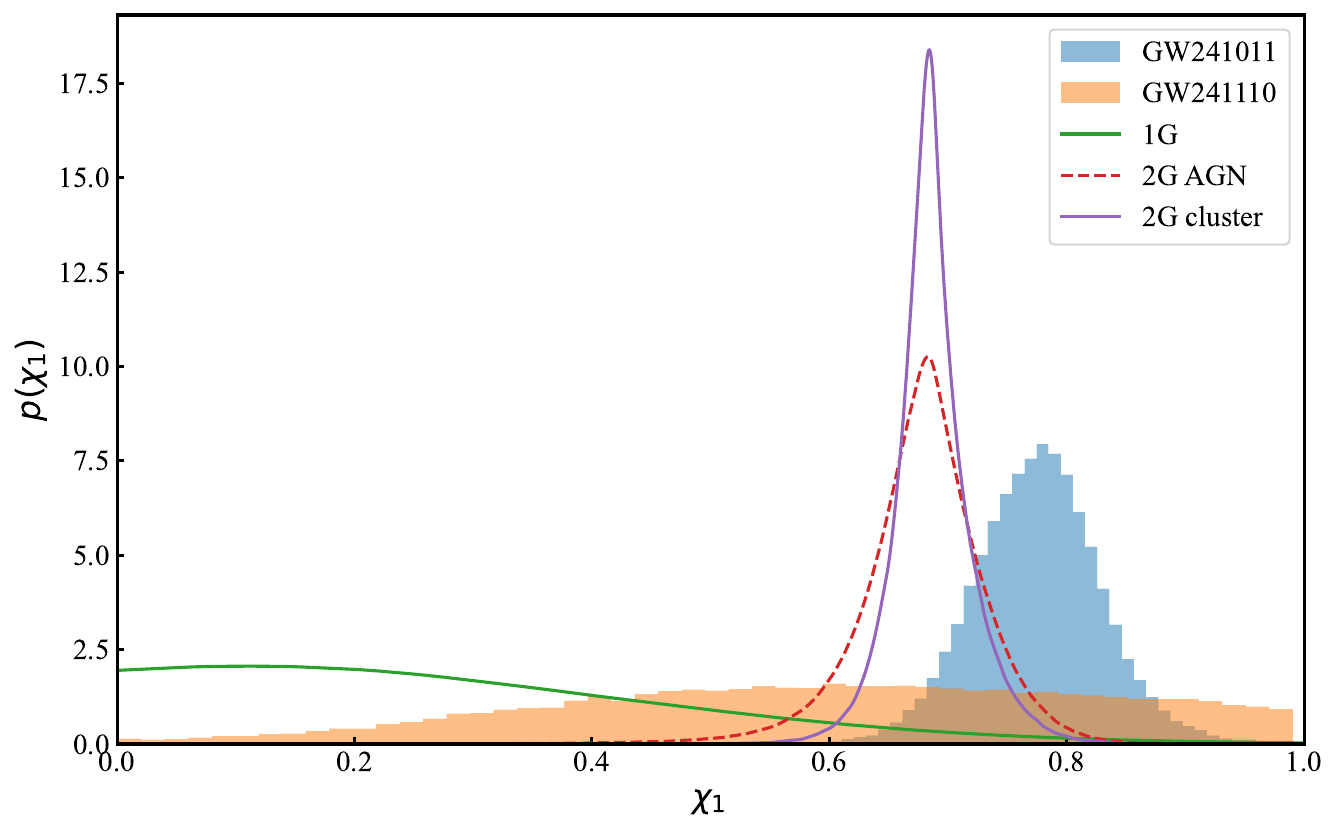}
    \includegraphics[width=0.48\textwidth]{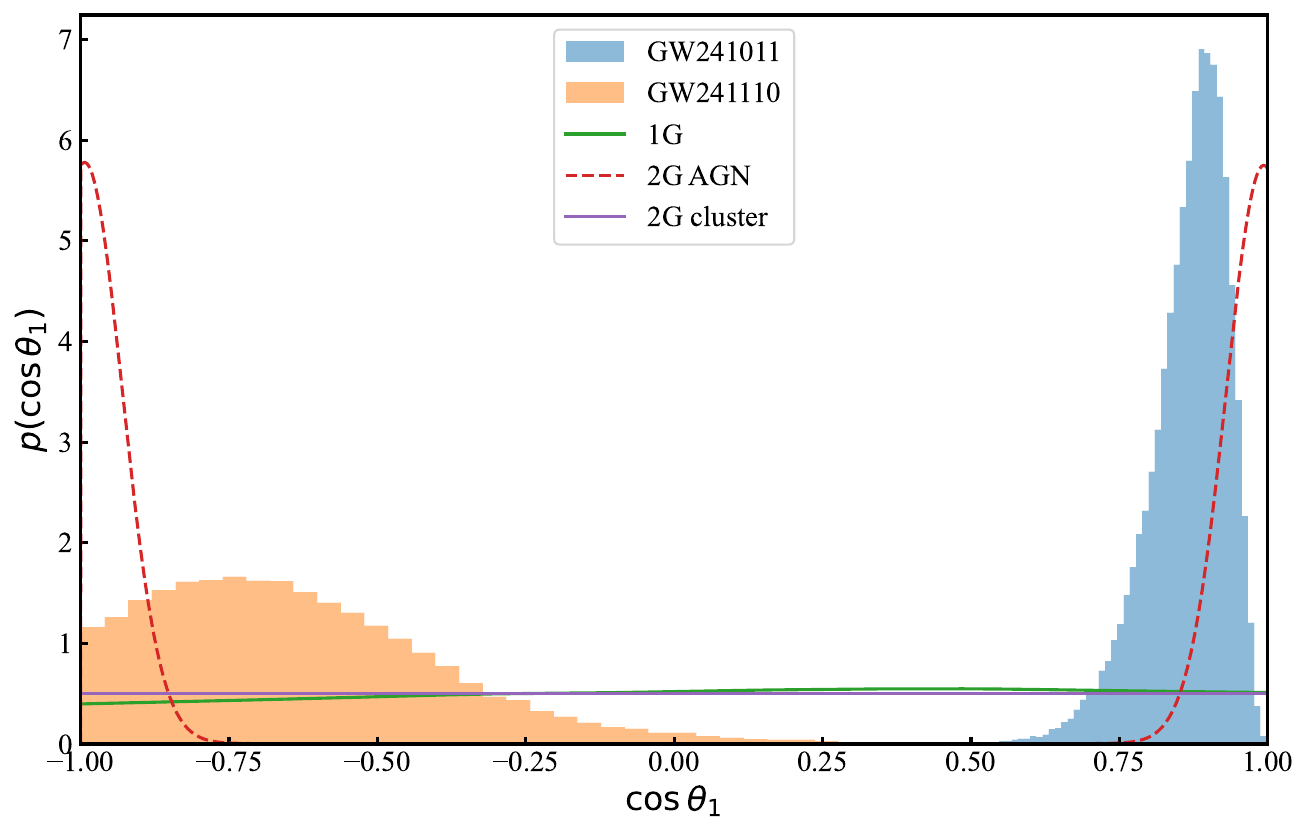}
  }
  \caption{Population prior distributions used in this work. The four panels show the distributions of the primary mass $m_1$, mass ratio $q$, primary spin magnitude $\chi_1$, and primary spin tilt $\cos\theta_1$, respectively. The population priors include the fiducial 1G+1G prior and the 2G+1G priors for the cluster and AGN disk. The posterior distributions of GW241011 and GW241110 are overlaid in each panel.}
  \label{fig:params_distribution}
\end{figure*}

Figure~\ref{fig:params_distribution} shows the resulting distributions of $m_1$, $q$, $\chi_1$, and $\cos\theta_1$ for the fiducial 1G+1G population, the 2G+1G population in the star cluster, and the 2G+1G population in the AGN disk. For comparison, we also show the corresponding posterior distributions of GW241011 and GW241110. 

In our 2G+1G hypotheses, we assumed that the primary BH is a 2G BH, while the secondary BH is a 1G BH. An alternative configuration, in which the secondary BH is a 2G remnant and the primary BH is first-generation, is also possible in principle. However, the inferred mass ratios for these two events are relatively low, making such a configuration less likely, and we therefore do not include this in our analysis. 

This construction can be applied recursively to generate priors for higher-generation BBH mergers, similar to the hierarchical growth model in \citet{liComparingHierarchicalBlack2023}. For example, retained 2G BHs can merge with 1G companions to produce 3G BHs. The retained 3G remnants can then be paired with 1G companions to construct the population prior for the 3G+1G hypotheses ($\mathcal{H}_{\rm 3G+1G}$).

\section{Results\label{sec:results}}

\subsection{Evidence for Hierarchical Mergers}

We first compare the fiducial 1G+1G hypothesis with the 2G+1G hypotheses in the cluster and the AGN disk channels. The resulting Bayes factors are summarized in Table~\ref{tab:result_2G1G_GW241011} and Table~\ref{tab:result_2G1G_GW241110} for GW241011 and GW241110, respectively.

\begin{deluxetable}{ccc}[htb!]
  \tablecaption{Bayes factors for GW241011 under the 2G+1G hypotheses.}
  \tablehead{
    \colhead{waveform} & $\ln\mathcal{B}^{\rm 2G+1G, cluster}_{\rm 1G+1G}$ & $\ln\mathcal{B}^{\rm 2G+1G, AGN}_{\rm 1G+1G}$
  }
  \label{tab:result_2G1G_GW241011}
  \startdata
  \texttt{IMRPhenomXO4a} & 6.5 & 7.1  \\
  \texttt{SEOBNRv5PHM} & 7.5 & 8.5  \\
  \texttt{IMRPhenomXPHM-SpinTaylor} & 7.3 & 8.6  \\
  \enddata
\end{deluxetable}

\begin{deluxetable}{ccc}[htb!]
  \tablecaption{Bayes factors for GW241110 under the 2G+1G hypotheses.}
  \tablehead{
    \colhead{waveform} & $\ln\mathcal{B}^{\rm 2G+1G, cluster}_{\rm 1G+1G}$ & $\ln\mathcal{B}^{\rm 2G+1G, AGN}_{\rm 1G+1G}$
  }
  \label{tab:result_2G1G_GW241110}
  \startdata
  \texttt{IMRPhenomXO4a} & 3.0 & 3.7 \\
  \texttt{SEOBNRv5PHM} & 3.6 & 4.5 \\
  \texttt{IMRPhenomXPHM-SpinTaylor} & 3.6 & 4.4 \\
  \enddata
\end{deluxetable}

For both events, the Bayes factors are positive for all available waveform models, indicating that the 2G+1G hypotheses are favored over the fiducial 1G+1G hypothesis. This preference can be understood from the parameter distributions shown in Figure~\ref{fig:params_distribution}. Compared with the fiducial 1G+1G prior, the 2G+1G priors extend to larger primary masses, have a mass ratio distribution peaking around 0.5 and favor primary spin magnitudes around 0.7, as expected for black holes produced by previous mergers \citep{gerosaHierarchicalMergersStellarmass2021}. The posterior distributions of GW241011 and GW241110 overlap more strongly with these hierarchical-merger priors than with the 1G+1G prior, leading to positive Bayes factors for the 2G+1G hypotheses.

The strength of the hierarchical-merger preference differs between the two events. GW241011 shows larger Bayes factors than GW241110, with $\ln\mathcal{B} > 5$ for all waveform models and both channels, whereas GW241110 shows more moderate support, with $\ln\mathcal{B}\simeq 3$--$5$. As shown in Figure~\ref{fig:params_distribution}, both events have large primary masses and small mass ratios, which are broadly consistent with a hierarchical merger interpretation. In addition, GW241011 has a higher and more concentrated primary spin magnitude posterior, making it more difficult to accommodate with the fiducial 1G+1G prior. This stronger contrast with the first-generation naturally leads to larger Bayes factors for GW241011.

The main difference between the AGN disk and star cluster channels arises from their predicted spin tilt distributions. As shown in Figure~\ref{fig:params_distribution}, the primary spin tilt posteriors of both GW241011 and GW241110 deviate from a purely isotropic distribution, with GW241011 favoring an aligned primary spin and GW241110 favoring an anti-aligned configuration. These features are more naturally accommodated by the AGN disk 2G+1G prior than by the cluster prior.

\subsection{Could GW241011 be a 3G+1G merger?}

As shown in Figure~\ref{fig:params_distribution}, although the properties of GW241011 are more consistent with a 2G+1G origin than with a 1G+1G origin, its posterior distributions still exhibit noticeable offsets from the typical 2G+1G population. In particular, the mass ratio of GW241011 is concentrated around $1/3$, lower than the typical value of $1/2$ predicted for 2G+1G binaries. Its primary spin magnitude is centered around 0.8, slightly higher than the characteristic value of 0.7 expected for second-generation BHs. A 3G+1G merger, involving a more massive primary, may naturally produce a more asymmetric binary and could potentially better accommodate these properties. This motivates us to investigate whether GW241011 could have originated from a 3G+1G merger.

\begin{figure*}[htb!]

  \gridline{
    \includegraphics[width=0.48\textwidth]{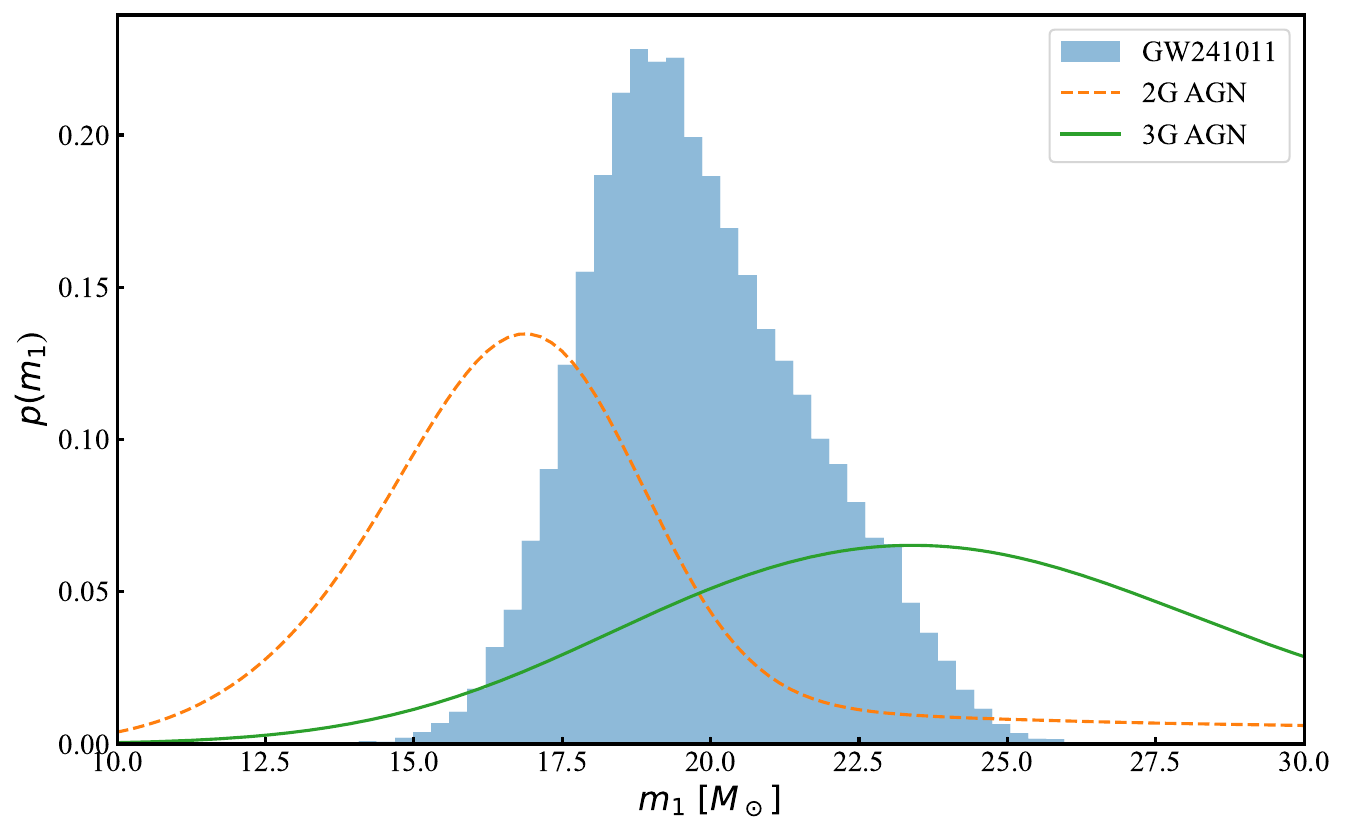}
    \includegraphics[width=0.48\textwidth]{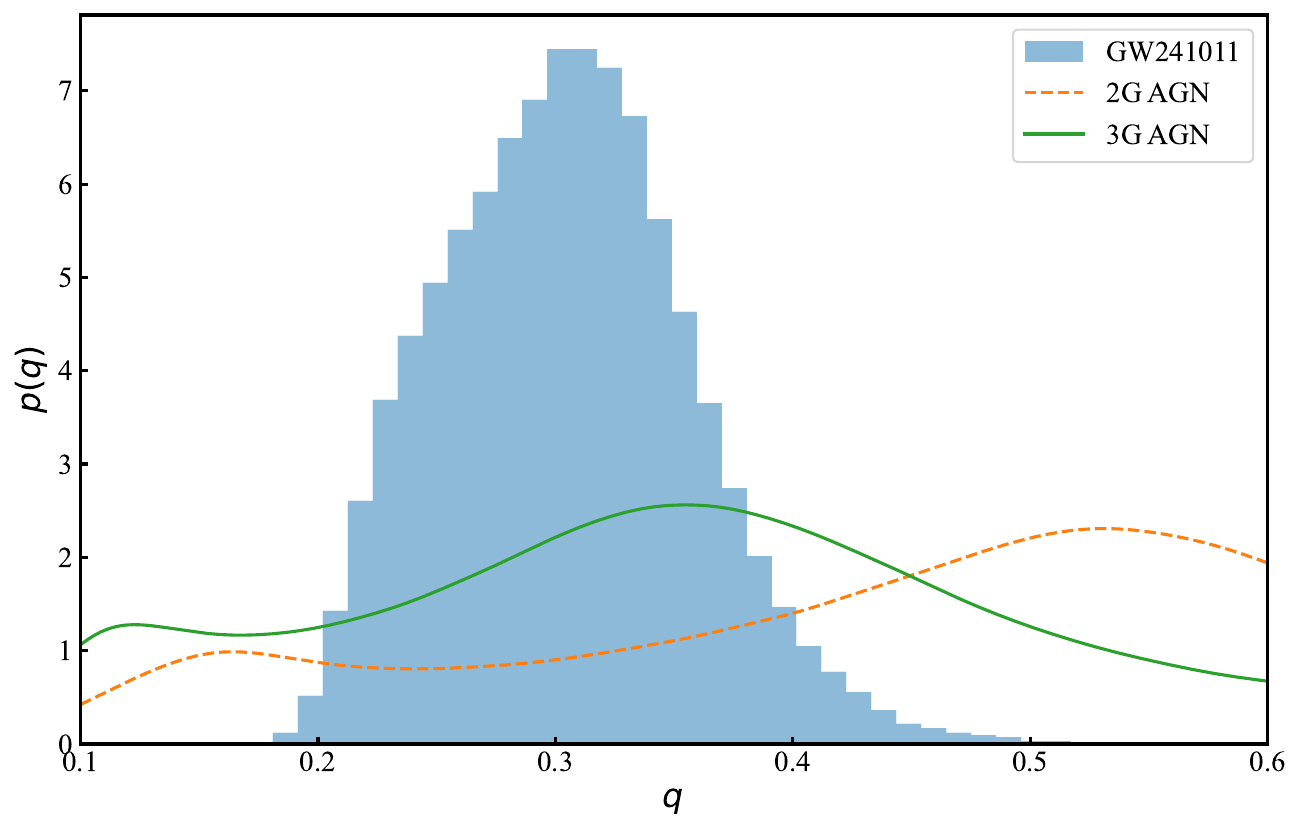}
  }
  \gridline{
    \includegraphics[width=0.48\textwidth]{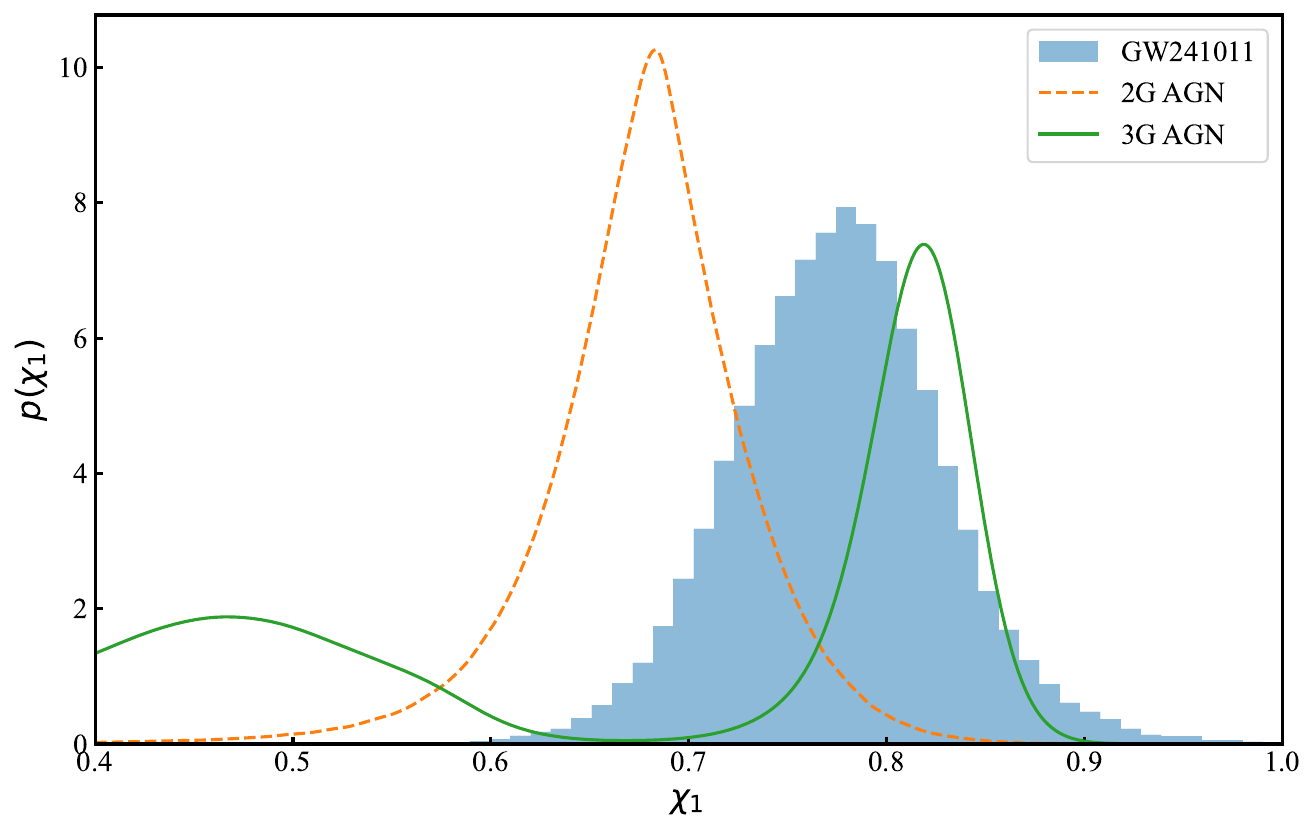}
    \includegraphics[width=0.48\textwidth]{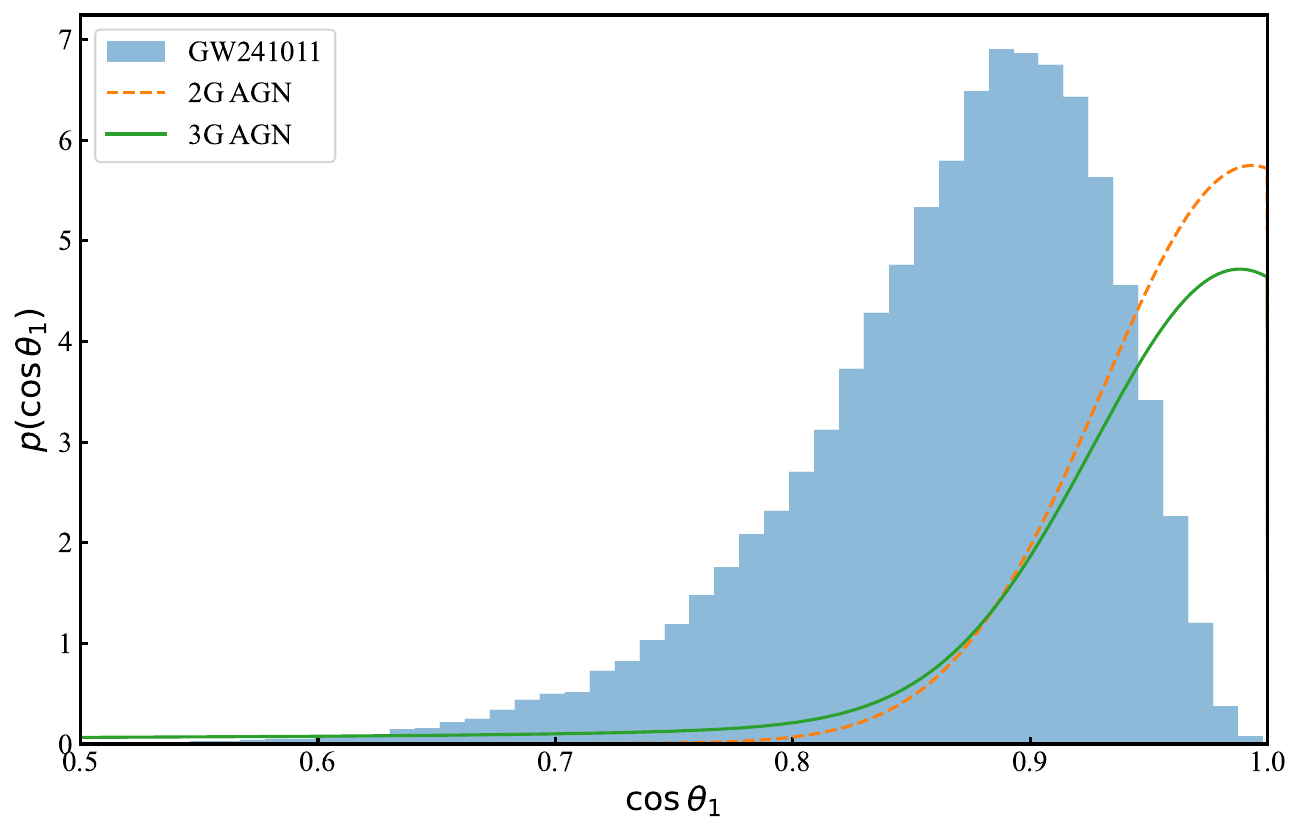}
  }
  \caption{Similar to Figure~\ref{fig:params_distribution}. The four panels show the distributions of the primary mass $m_1$, mass ratio $q$, primary spin magnitude $\chi_1$, and primary spin tilt $\cos\theta_1$, respectively.  The population priors include the 2G+1G and 3G+1G priors in the AGN disk environment. The posterior distribution of GW241011 is overlaid in each panel. }
  \label{fig:params_3g}
\end{figure*}

Considering that the formation of a 3G black hole requires two successive mergers, the rate of a 3G+1G merger in star clusters is expected to be strongly suppressed, because merger remnants may be kicked out of the host environment. Therefore, we only consider the AGN disk environment, where repeated mergers and the retention of higher-generation remnants are more readily achieved. The parameter distributions of a 3G+1G merger in AGN disk are shown in Figure~\ref{fig:params_3g}, together with the posterior distribution of GW241011.

As expected, the 3G+1G population is shifted toward larger primary masses and smaller mass ratios compared with the 2G+1G population. This is a natural consequence of replacing the second-generation primary with a more massive third-generation remnant while keeping the companion drawn from the first-generation population. The primary spin magnitude shows a main peak at a higher value, closer to the posterior of GW241011. A weaker low-spin component appears at $\chi_1 \sim 0.5$, mainly produced by mergers in which the spin of the higher-generation progenitor is anti-aligned with the orbital angular momentum, partially reducing the final remnant spin. The spin tilt distribution remains broadly similar to that of the 2G+1G population, but becomes slightly broader. This broadening occurs because the 2G progenitor has a relatively large spin, and its orientation can have a non-negligible effect on the direction of the final remnant spin. In contrast, in the formation of the 2G black holes, the spins of the two 1G progenitors are typically smaller, so the remnant spin direction is more strongly dominated by the orbital angular momentum.

We calculate the Bayes factors for the 3G+1G AGN hypothesis relative to both the fiducial 1G+1G and 2G+1G AGN hypotheses for GW241011, and the results are listed in Table~\ref{tab:result_3G1G_GW241011}. We find that the 3G+1G AGN hypothesis is clearly favored over the 1G+1G interpretation. However, its evidence relative to the 2G+1G AGN hypothesis is weak and waveform dependent, with $\ln\mathcal{B}^{\rm 3G+1G,AGN}_{\rm 2G+1G,AGN}$ ranging from $-1.9$ to $1.1$. Therefore, the data do not robustly prefer a 3G+1G origin over a 2G+1G origin. In addition, since the formation of a 3G BH requires two previous mergers, 3G BHs are expected to be less common than 2G BHs. This additional rate suppression would further weaken the astrophysical plausibility of the 3G+1G interpretation. We therefore conclude that GW241011 does not require a 3G+1G origin.

\begin{deluxetable}{ccc}[htb!]
  \tablecaption{Bayes factors for GW241011 under the 3G+1G AGN hypothesis.}
  \tablehead{
    \colhead{waveform} & $\ln\mathcal{B}^{\rm 3G+1G, AGN}_{\rm 1G+1G}$ & $\ln\mathcal{B}^{\rm 3G+1G, AGN}_{\rm 2G+1G, AGN}$
  }
  \label{tab:result_3G1G_GW241011}
  \startdata
  \texttt{IMRPhenomXO4a} & 8.2 & 1.1 \\
  \texttt{SEOBNRv5PHM} & 9.1 & 0.6 \\
  \texttt{IMRPhenomXPHM-SpinTaylor} & 6.7 & -1.9 \\
  \enddata
\end{deluxetable}

\section{Searching for Electromagnetic Counterpart\label{sec:search}}

Our analysis suggests that GW241011 and GW241110 are promising hierarchical merger candidates, with a mild preference for the AGN disk channel over the star cluster channel. If these mergers occurred in AGN disks, interactions between the binary or merger remnant and the surrounding gas could produce observable EM emission. We therefore perform an optical search for candidate EM counterparts associated with GW241011 and GW241110.

\begin{figure*}[htb!]

  \gridline{
    \includegraphics[width=0.48\textwidth]{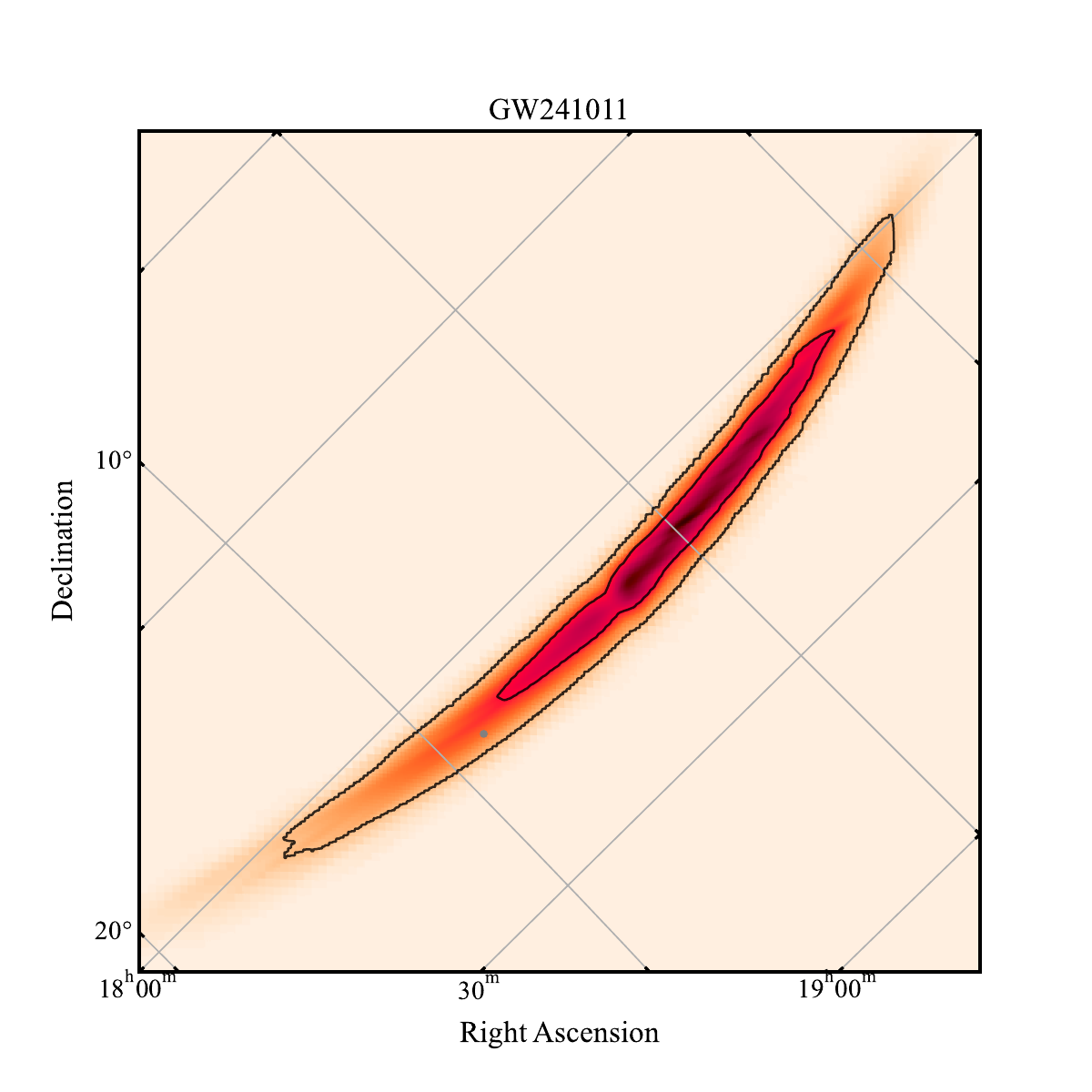}
    \includegraphics[width=0.48\textwidth]{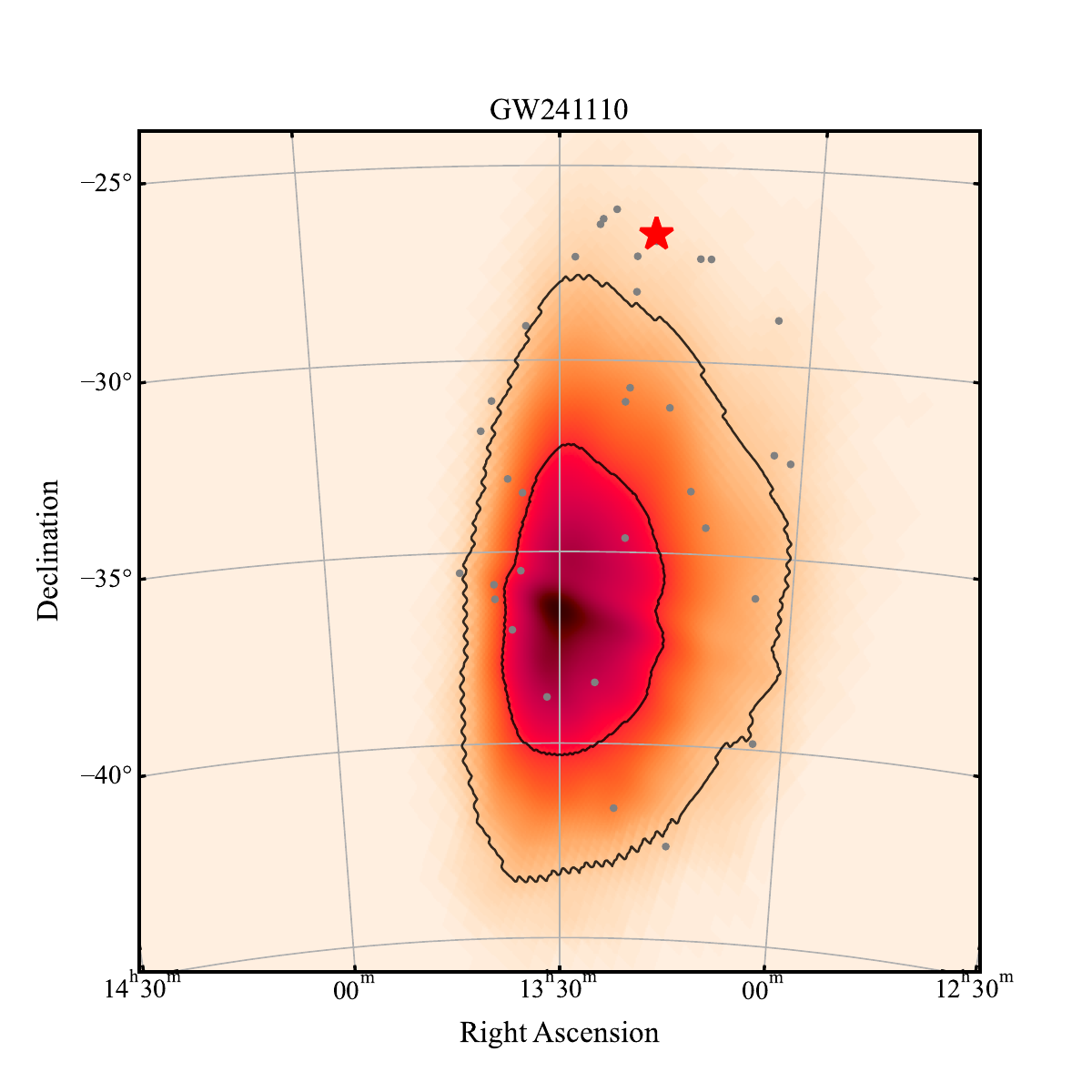}
  }
  \caption{Skymaps for GW241011 and GW241110 constructed from the \texttt{Mixed} posterior samples. Contours indicate the 50\% and 90\% credible localization regions. Gray points denote AGNs located within the 95\% three-dimensional credible localization volume of each event. The red star marks the AGN with a candidate flare found in our search. 
}
  \label{fig:skymap}
\end{figure*}

For the search, we adopt the skymaps constructed from the \texttt{Mixed} posterior samples for both events. We construct a parent AGN sample by combining confirmed AGNs with available redshift measurements from the Million Quasars catalog \citep{fleschMillionQuasarsMilliquas2023}, the Dark Energy Spectroscopic Instrument (DESI) survey \citep{collaborationDataRelease12026}, and the Large Sky Area Multi-Object Fiber Spectroscopic Telescope (LAMOST) survey \citep{aiLargeSkyArea2016,dongLargeSkyArea2018,yaoLargeSkyArea2019,jinLargeSkyArea2023,lyuLargeSkyArea2026}. We then crossmatch this AGN sample with the GW skymaps and select sources located within 95\% three-dimensional credible localization volume of each event. This procedure yields one AGN for GW241011 and 36 AGNs for GW241110. The skymaps and the locations of these selected AGNs are shown in Figure~\ref{fig:skymap}.

For these cross-matched AGNs, we collect archival forced-photometry light curves from the Zwicky Transient Facility \citep[ZTF,][]{masciNewForcedPhotometry2023} and the Asteroid Terrestrial-impact Last Alert System \citep[ATLAS,][]{smithDesignOperationATLAS2020}. We use the ZTF $g$- and $r$-band light curves and the ATLAS $c$- and $o$-band light curves to characterize the variability of each source. The data are reduced following the recommended procedures \footnote{ZTF document: \url{https://irsa.ipac.caltech.edu/data/ZTF/docs/ztf_forced_photometry.pdf},\\ ATLAS FAQ: \url{https://fallingstar-data.com/forcedphot/faq }}.

To find the flare candidates in the optical light curves, we apply the Bayesian Blocks algorithm \citep{scargleSTUDIESASTRONOMICALTIME2013} to the light curves in each band. This method partitions an unevenly sampled light curve into a set of time intervals within which the flux is statistically consistent with being constant, while allowing changes in the flux level between adjacent blocks. We then search for post-trigger blocks with flux levels higher than both the adjacent blocks and the baseline flux level, following the selection of potential flares used in \citet{heSystematicSearchActive2025}. 

\begin{figure*}[htb!]
  \includegraphics[width=\textwidth]{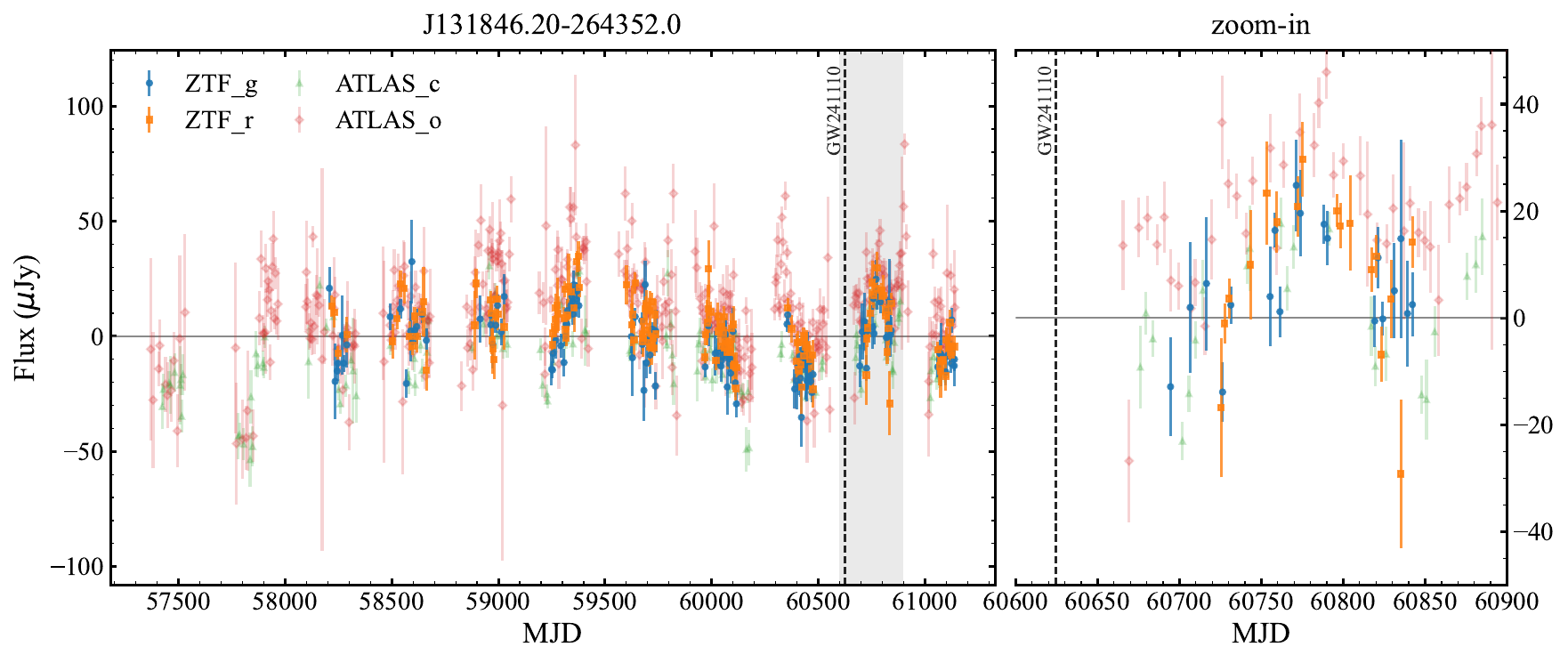}
  \caption{Forced-photometry light curves of J131846.20-264352.0 from ZTF and ATLAS. The vertical dashed line marks the trigger time of GW241110. The shaded region indicates the time interval of the possible flare identified by the Bayesian Blocks analysis. The right panel shows a zoom-in view of the flare interval.}
  \label{fig:lightcurve}
\end{figure*}

Using this method, we find that one AGN, J131846.20-264352.0, located within the localization volume of GW241110, shows post-trigger flare-like variability. This source has a spectroscopic redshift of $z=0.1465$ from DESI and lies at the 93\% 3D contour of GW241110. However, inspection of its long-term light curve (Figure~\ref{fig:lightcurve}) shows that this brightening is not statistically compelling compared with the intrinsic variability of the source. Therefore, we regard J131846.20-264352.0 only as a tentative flare candidate rather than a confirmed EM counterpart.

Continued monitoring of this source is necessary to further assess its nature. In the AGN disk scenario, the kicked merger remnant may remain bound to the central supermassive black hole and reencounter the disk, potentially producing recurrent flares \citep{grahamCandidateElectromagneticCounterpart2020}. A second flare, if detected, would provide a useful clue to the possible merger location in the AGN disk. In addition, off-center emission from a kicked remnant may lead to asymmetric broad emission lines \citep{mckernanRampressureStrippingKicked2019}. Therefore, future spectroscopic follow-up during any recurrent flare would be crucial for distinguishing a possible BBH-related flare from intrinsic AGN variability.

The absence of a compelling EM counterpart in our search is not unexpected, given several observational and physical limitations. For GW241011, the localization region lies close to the Galactic plane, where AGN samples are highly incomplete because of severe stellar contamination and dust extinction. Therefore, the number of known AGNs available for cross-matching in this region is likely underestimated. For GW241110, a large fraction of the localization region falls below the southern declination limit of the ZTF footprint, $\mathrm{Dec.}\sim -30^\circ$ \citep{bellmZwickyTransientFacility2019}. As a result, many of the candidate AGNs in this region have only ATLAS coverage. Compared with ZTF, ATLAS is shallower and typically has larger photometric uncertainties, making it more difficult to identify faint flares. 

In addition, the remnant black holes of GW241011 and GW241110 are not exceptionally massive compared with the most massive BBH mergers, such as GW231123 \citep{abacGW231123BinaryBlack2025a,heSearchingElectromagneticCounterpart2025}. Moreover, their inferred kick velocities are relatively large. The kick velocity is estimated to be $974^{+555}_{-466} \, \mathrm {km \,s^{-1}}$ for GW241011 and $394^{+582}_{-207}\,\mathrm{km \,s^{-1}}$ for GW241110 \citep{islamInferenceRecoilKicks2026}, with median values above the typical scale of $\sim 200 \,\mathrm{km \, s^{-1}}$. Since the Bondi-Hoyle-Lyttleton luminosity scales as $L_{\rm BHL} \propto M_{\rm BBH}^2 / v_{\rm k}^3$ \citep{grahamCandidateElectromagneticCounterpart2020}, such properties may result in relatively faint emission for the mergers. Detectable optical flares from these events may require particularly favorable disk conditions, such as a high local gas density. As a result, only a restricted region of parameter space is likely to produce flares bright enough to stand out above the intrinsic AGN variability and be detected by current optical surveys. 

\section{Discussion and Conclusions\label{sec:conclusion}}

We have investigated whether GW241011 and GW241110 can be explained as hierarchical BBH mergers. Both events have rapidly spinning primary BHs and unequal component masses, making them natural candidates for systems in which the primary BH is the remnant of a previous BBH merger \citep{gerosaHierarchicalMergersStellarmass2021,abacGW241011GW241110Exploring2025}. Using the posterior samples released by LVK, we compared the fiducial first-generation hypothesis with hierarchical merger hypotheses in star clusters and AGN disks.

Both events favor a 2G+1G interpretation over the fiducial 1G+1G population. For GW241011, the support is strong across all available waveform models, with $\ln\mathcal{B}\simeq6.5-7.5$ for the star cluster channel and $\ln\mathcal{B}\simeq7.1-8.6$ for the AGN disk channel. For GW241110, the support is more moderate but remains positive, with $\ln\mathcal{B}\simeq3.0-3.6$ for the star cluster channel and $\ln\mathcal{B}\simeq3.7-4.5$ for the AGN disk channel. These results are broadly consistent with recent independent studies that identify GW241011 and GW241110 as possible hierarchical merger candidates \citep{tongSubpopulationLowmassSpinning2025,liGW241011GW241110Hints2025}.

The comparison between cluster and AGN disk environment is more subtle. The AGN disk 2G+1G hypothesis gives slightly larger Bayes factors for both events, mainly because GW241011 favors a positive primary spin projection while GW241110 favors a negative primary spin projection, both of which are more naturally matched by the aligned or anti-aligned spin geometry in the AGN disk environment \citep{yangHierarchicalBlackHole2019}. However, the evidence difference between the two environments is not large enough to claim a decisive environment classification. We also tested a 3G+1G hypothesis for GW241011. Although this hypothesis is favored over 1G+1G, its evidence relative to 2G+1G is weak and waveform dependent, so the present data do not require a third-generation origin.

Motivated by the mild AGN disk preference, we searched for optical counterpart candidates by corssmatching known AGNs with the GW three-dimensional localization volumes and inspecting their ZTF and ATLAS forced-photometry light curves. We find one AGN that shows a possible brightening roughly 100 days after the GW trigger, but its long-term light curve variability does not provide compelling evidence for a physical association. Continued photometric monitoring and spectroscopic follow-up will be valuable for assessing the nature of this source.

Several limitations should be kept in mind when interpreting these results. The Bayes factors presented here measure the consistency of individual events with the adopted formation priors, rather than the full astrophysical odds of the competing channels, since relative merger rates and selection effects are not included. We also do not include alternative isolated-binary channels, such as the recently proposed mass-ratio reversal scenario for GW241011-like systems \citep{huMassRatioReversalAlternative2026}. In addition, our population, retention, and spin-orientation prescriptions are simplified. More realistic population models and population-level analyses will be needed to place stronger constraints on the origin of these events.

\begin{acknowledgements}
This work is supported by the National Natural Science Foundation of China (grant Nos. 12325301, 12273035 and 12405075), Strategic Priority Research Program of the Chinese Academy of Science (grant No. XDB0550300), the National Key R\&D Program of China (grant Nos. 2021YFC2203102, 2022YFC2204602, and 2024YFC2207500), and Cyrus Chun Ying Tang Foundations. 

The ZTF forced-photometry service was funded under the Heising-Simons Foundation grant \#12540303 (PI: Graham). This work has made use of data from the Asteroid Terrestrial-impact Last Alert System (ATLAS) project. The Asteroid Terrestrial-impact Last Alert System (ATLAS) project is primarily funded to search for near earth asteroids through NASA grants NN12AR55G, 80NSSC18K0284, and 80NSSC18K1575; byproducts of the NEO search include images and catalogs from the survey area. This work was partially funded by Kepler/K2 grant J1944/80NSSC19K0112 and HST GO-15889, and STFC grants ST/T000198/1 and ST/S006109/1. The ATLAS science products have been made possible through the contributions of the University of Hawaii Institute for Astronomy, the Queen’s University Belfast, the Space Telescope Science Institute, the South African Astronomical Observatory, and The Millennium Institute of Astrophysics (MAS), Chile.

\end{acknowledgements}

\software{numpy \citep{harrisArrayProgrammingNumPy2020}, matplotlib \citep{hunterMatplotlib2DGraphics2007}, bilby \citep{ashtonBilbyUserfriendlyBayesian2019}, dynesty \citep{speagleDynestyDynamicNested2020}, figaro \citep{rinaldiFIGAROHierarchicalNonparametric2024}, ligo.skymap \citep{singerRapidBayesianPosition2016}, astropy \citep{astropycollaborationAstropyCommunityPython2013,astropycollaborationAstropyProjectBuilding2018,astropycollaborationAstropyProjectSustaining2022}, pandas \citep{mckinneyDataStructuresStatistical2010}
}
\bibliography{main}

@article{abacGW231123BinaryBlack2025a,
  title = {{{GW231123}}: {{A Binary Black Hole Merger}} with {{Total Mass}} 190--265 {{M}}{$\odot$}},
  shorttitle = {{{GW231123}}},
  author = {Abac, A. G. and Abouelfettouh, I. and Acernese, F. and Ackley, K. and Adamcewicz, C. and Adhicary, S. and Adhikari, D. and Adhikari, N. and Adhikari, R. X. and Adkins, V. K. and Afroz, S. and Agapito, A. and Agarwal, D. and Agathos, M. and Aggarwal, N. and Aggarwal, S. and Aguiar, O. D. and Ahrend, I.-L. and Aiello, L. and Ain, A. and Ajith, P. and Akutsu, T. and Albanesi, S. and Ali, W. and {Al-Kershi}, S. and All{\'e}n{\'e}, C. and Allocca, A. and {Al-Shammari}, S. and Altin, P. A. and {Alvarez-Lopez}, S. and Amar, W. and Amarasinghe, O. and Amato, A. and Amicucci, F. and Amra, C. and Ananyeva, A. and Anderson, S. B. and Anderson, W. G. and Andia, M. and Ando, M. and {Andr{\'e}s-Carcasona}, M. and Andri{\'c}, T. and Anglin, J. and Ansoldi, S. and Antelis, J. M. and Antier, S. and Aoumi, M. and Appavuravther, E. Z. and Appert, S. and Apple, S. K. and Arai, K. and Araujo Alvarez, C. and Araya, A. and Araya, M. C. and Arca Sedda, M. and Areeda, J. S. and Aritomi, N. and Armato, F. and Armstrong, S. and Arnaud, N. and Arogeti, M. and Aronson, S. M. and Arun, K. G. and Ashton, G. and Aso, Y. and Asprea, L. and Assiduo, M. and {Assis de Souza Melo}, S. and Aston, S. M. and Astone, P. and Attadio, F. and Aubin, F. and AultONeal, K. and Avallone, G. and Avila, E. A. and Babak, S. and Badger, C. and Bae, S. and Bagnasco, S. and Baiotti, L. and Bajpai, R. and Baka, T. and Baker, A. M. and Baker, K. A. and Baker, T. and Baldi, G. and Baldicchi, N. and Ball, M. and Ballardin, G. and Ballmer, S. W. and Banagiri, S. and Banerjee, B. and Bankar, D. and Baptiste, T. M. and Baral, P. and Baratti, M. and Barayoga, J. C. and Barish, B. C. and Barker, D. and Barman, N. and Barneo, P. and Barone, F. and Barr, B. and Barsotti, L. and Barsuglia, M. and Barta, D. and Bartoletti, A. M. and Barton, M. A. and Bartos, I. and Basalaev, A. and Bassiri, R. and Basti, A. and Bawaj, M. and Baxi, P. and Bayley, J. C. and Baylor, A. C. and Baynard II, P. A. and Bazzan, M. and Bedakihale, V. M. and Beirnaert, F. and Bejger, M. and Belardinelli, D. and Bell, A. S. and Bellie, D. S. and Bellizzi, L. and Benoit, W. and Bentara, I. and Bentley, J. D. and Ben Yaala, M. and Bera, S. and Bergamin, F. and Berger, B. K. and Bernuzzi, S. and Beroiz, M. and Berry, C. P. L. and Bersanetti, D. and Bertheas, T. and Bertolini, A. and Betzwieser, J. and Beveridge, D. and Bevilacqua, G. and Bevins, N. and Bhandare, R. and Bhatt, R. and Bhattacharjee, D. and Bhattacharyya, S. and Bhaumik, S. and Bhagwat, S. and Biancalana, V. and Bianchi, A. and Bilenko, I. A. and Billingsley, G. and Binetti, A. and Bini, S. and Binu, C. and Biot, S. and Birnholtz, O. and Biscoveanu, S. and Bisht, A. and Bitossi, M. and Bizouard, M.-A. and Blaber, S. and Blackburn, J. K. and Blagg, L. A. and Blair, C. D. and Blair, D. G. and Bode, N. and Boettner, N. and Boileau, G. and Boldrini, M. and Bolingbroke, G. N. and Bolliand, A. and Bonavena, L. D. and Bondarescu, R. and Bondu, F. and Bonilla, E. and Bonilla, M. S. and Bonino, A. and Bonnand, R. and Borchers, A. and Borhanian, S. and Boschi, V. and Bose, S. and Bossilkov, V. and Bothra, Y. and Boudon, A. and Bourg, L. and Boyle, M. and Bozzi, A. and Bradaschia, C. and Brady, P. R. and Branch, A. and Branchesi, M. and Braun, I. and Briant, T. and Brillet, A. and Brinkmann, M. and Brockill, P. and Brockmueller, E. and Brooks, A. F. and Brown, B. C. and Brown, D. D. and Brozzetti, M. L. and Brunett, S. and Bruno, G. and Bruntz, R. and Bryant, J. and Bu, Y. and Bucci, F. and Buchanan, J. and Bulashenko, O. and Bulik, T. and Bulten, H. J. and Buonanno, A. and Burtnyk, K. and Buscicchio, R. and Buskulic, D. and Buy, C. and Byer, R. L. and Cabourn Davies, G. S. and Cabrita, R. and {C{\'a}ceres-Barbosa}, V. and Cadonati, L. and Cagnoli, G. and Cahillane, C. and Calafat, A. and Calder{\'o}n Bustillo, J. and Callister, T. A. and Calloni, E. and Callos, S. R. and Canepa, M. and Caneva Santoro, G. and Cannon, K. C. and Cao, H. and Capistran, L. A. and Capocasa, E. and Capote, E. and Capurri, G. and Carapella, G. and Carbognani, F. and Carlassara, M. and Carlin, J. B. and Carlson, T. K. and Carney, M. F. and Carpinelli, M. and Carrillo, G. and Carter, J. J. and Carullo, G. and {Casallas-Lagos}, A. and Casanueva Diaz, J. and Casentini, C. and {Castro-Lucas}, S. Y. and Caudill, S. and Cavagli{\`a}, M. and Cavalieri, R. and Ceja, A. and Cella, G. and {Cerd{\'a}-Dur{\'a}n}, P. and Cesarini, E. and Chabbra, N. and Chaibi, W. and Chakraborty, A. and Chakraborty, P. and Chakraborty, S. and Chalathadka Subrahmanya, S. and Chan, J. C. L. and Chan, M. and Chandra, K. and Chang, K. and Chao, S. and Charlton, P. and {Chassande-Mottin}, E. and Chatterjee, C. and Chatterjee, Debarati and Chatterjee, Deep and Chaturvedi, M. and Chaty, S. and Chatziioannou, K. and Chen, A. and Chen, A. H.-Y. and Chen, D. and Chen, H. and Chen, H. Y. and Chen, S. and Chen, Yanbei and Chen, Yitian and Cheng, H. P. and Chessa, P. and Cheung, H. T. and Cheung, S. Y. and Chiadini, F. and Chiarini, G. and Chiba, A. and Chincarini, A. and Chiofalo, M. L. and Chiummo, A. and Chou, C. and Choudhary, S. and Christensen, N. and Chua, S. S. Y. and Ciani, G. and Ciecielag, P. and Cie{\'s}lar, M. and Cifaldi, M. and Cirok, B. and Clara, F. and Clark, J. A. and Clarke, T. A. and Clearwater, P. and Clesse, S. and Cleva, F. and Coccia, E. and Codazzo, E. and Cohadon, P.-F. and Colace, S. and Colangeli, E. and Colleoni, M. and Collette, C. G. and Collins, J. and Colloms, S. and Colombo, A. and Compton, C. M. and Connolly, G. and Conti, L. and Corbitt, T. R. and {Cordero-Carri{\'o}n}, I. and Corezzi, S. and Cornish, N. J. and Coronado, I. and Corsi, A. and Cottingham, R. and Coughlin, M. W. and Couineaux, A. and Couvares, P. and Coward, D. M. and Coyne, R. and Cozzumbo, A. and Creighton, J. D. E. and Creighton, T. D. and Cremonese, P. and Crook, S. and Crouch, R. and Csizmazia, J. and Cudell, J. R. and Cullen, T. J. and Cumming, A. and Cuoco, E. and Cusinato, M. and Da Concei{\c c}{\~a}o, L. V. and Dal Canton, T. and Dal Pra, S. and D{\'a}lya, G. and D'Angelo, B. and Danilishin, S. and D'Antonio, S. and Danzmann, K. and Darroch, K. E. and Dartez, L. P. and Das, R. and Dasgupta, A. and Dattilo, V. and Daumas, A. and Davari, N. and Dave, I. and Davenport, A. and Davier, M. and Davies, T. F. and Davis, D. and Davis, L. and Davis, M. C. and Davis, P. and Daw, E. J. and Dax, M. and De Bolle, J. and Deenadayalan, M. and Degallaix, J. and De Laurentis, M. and De Lillo, F. and Della Torre, S. and Del Pozzo, W. and Demagny, A. and De Marco, F. and Demasi, G. and De Matteis, F. and Demos, N. and Dent, T. and Depasse, A. and DePergola, N. and De Pietri, R. and De Rosa, R. and De Rossi, C. and Desai, M. and DeSalvo, R. and DeSimone, A. and De Simone, R. and Dhani, A. and Diab, R. and D{\'i}az, M. C. and Di Cesare, M. and Dideron, G. and Dietrich, T. and Di Fiore, L. and Di Fronzo, C. and Di Giovanni, M. and Di Girolamo, T. and Diksha, D. and Ding, J. and Di Pace, S. and Di Palma, I. and Di Piero, D. and Di Renzo, F. and {Divyajyoti} and Dmitriev, A. and Docherty, J. P. and Doctor, Z. and Doerksen, N. and Dohmen, E. and Doke, A. and Domiciano De Souza, A. and D'Onofrio, L. and Donovan, F. and Dooley, K. L. and Dooney, T. and Doravari, S. and Dorosh, O. and Doyle, W. J. D. and Drago, M. and Driggers, J. C. and Dunn, L. and Dupletsa, U. and Duverne, P.-A. and D'Urso, D. and Dutta Roy, P. and Duval, H. and Dwyer, S. E. and Eassa, C. and Ebersold, M. and Eckhardt, T. and Eddolls, G. and Effler, A. and Eichholz, J. and Einsle, H. and Eisenmann, M. and Emma, M. and Endo, K. and Enficiaud, R. and Errico, L. and Espinosa, R. and Esposito, M. and Essick, R. C. and Estell{\'e}s, H. and Etzel, T. and Evans, M. and Evstafyeva, T. and Ewing, B. E. and Ezquiaga, J. M. and Fabrizi, F. and Fafone, V. and Fairhurst, S. and Farah, A. M. and Farr, B. and Farr, W. M. and Favaro, G. and Favata, M. and Fays, M. and Fazio, M. and Feicht, J. and Fejer, M. M. and Felicetti, R. and Fenyvesi, E. and Fernandes, J. and Fernandes, T. and Fernando, D. and Ferraiuolo, S. and Ferreira, T. A. and Fidecaro, F. and Figura, P. and Fiori, A. and Fiori, I. and Fisher, R. P. and Fittipaldi, R. and Fiumara, V. and Flaminio, R. and Fleischer, S. M. and Fleming, L. S. and Floden, E. and Fong, H. and Font, J. A. and {Fontinele-Nunes}, F. and Foo, C. and Fornal, B. and Franceschetti, K. and Franchini, N. and Frappez, F. and Frasca, S. and Frasconi, F. and Freed, J. P. and Frei, Z. and Freise, A. and Freitas, O. and Frey, R. and Frischhertz, W. and Fritschel, P. and Frolov, V. V. and Fronz{\'e}, G. G. and {Fuentes-Garcia}, M. and Fujii, S. and Fujimori, T. and Fulda, P. and Fyffe, M. and Gadre, B. and Gair, J. R. and Galaudage, S. and Galdi, V. and Gamba, R. and Gamboa, A. and Gamoji, S. and Ganapathy, D. and Ganguly, A. and Garaventa, B. and {Garc{\'i}a-Bellido}, J. and {Garc{\'i}a-Quir{\'o}s}, C. and Gardner, J. W. and Gardner, K. A. and Garg, S. and Gargiulo, J. and Garrido, X. and Garron, A. and Garufi, F. and Garver, P. A. and Gasbarra, C. and Gateley, B. and Gautier, F. and Gayathri, V. and Gayer, T. and Gemme, G. and Gennai, A. and Gennari, V. and George, J. and George, R. and Gerberding, O. and Gergely, L. and Ghosh, Sayantan and Ghosh, Shaon and Ghosh, Shrobana and Ghosh, Suprovo and Ghosh, Tathagata and Giaime, J. A. and Giardina, K. D. and Gibson, D. R. and Gier, C. and Gkaitatzis, S. and Glanzer, J. and Glotin, F. and Godfrey, J. and Godley, R. V. and Godwin, P. and Goettel, A. S. and Goetz, E. and Golomb, J. and Gomez Lopez, S. and Goncharov, B. and Gonz{\'a}lez, G. and Goodarzi, P. and Goode, S. and Gosselin, M. and Gouaty, R. and Gould, D. W. and Govorkova, K. and Grado, A. and Graham, V. and Granados, A. E. and Granata, M. and Granata, V. and Gras, S. and Grassia, P. and Graves, J. and Gray, C. and Gray, R. and Greco, G. and Green, A. C. and Green, L. and Green, S. M. and Green, S. R. and Greenberg, C. and Gretarsson, A. M. and Griffin, H. K. and Griffith, D. and Griggs, H. L. and Grignani, G. and Grimaud, C. and Grote, H. and Grunewald, S. and Guerra, D. and Guetta, D. and Guidi, G. M. and Guimaraes, A. R. and Gulati, H. K. and Gulminelli, F. and Guo, H. and Guo, W. and Guo, Y. and Gupta, Anuradha and Gupta, I. and Gupta, N. C. and Gupta, S. K. and Gupta, V. and Gupte, N. and Gurs, J. and Gutierrez, N. and Guttman, N. and Guzman, F. and Haba, D. and Haberland, M. and Haino, S. and Hall, E. D. and Hamilton, E. Z. and Hammond, G. and Haney, M. and Hanks, J. and Hanna, C. and Hannam, M. D. and Hanselman, A. G. and Hansen, H. and Hanson, J. and Hanumasagar, S. and Harada, R. and Hardison, A. R. and Harikumar, S. and Haris, K. and {Harley-Trochimczyk}, I. and Harmark, T. and Harms, J. and Harry, G. M. and Harry, I. W. and Hart, J. and Haskell, B. and Haster, C. J. and Haughian, K. and Hayakawa, H. and Hayama, K. and Heintze, M. C. and Heinze, J. and Heinzel, J. and Heitmann, H. and Hellman, F. and {Helmling-Cornell}, A. F. and Hemming, G. and {Henderson-Sapir}, O. and Hendry, M. and Heng, I. S. and Hennig, M. H. and Henshaw, C. and Heurs, M. and Hewitt, A. L. and Heynen, J. and Heyns, J. and Higginbotham, S. and Hild, S. and Hill, S. and Himemoto, Y. and Hirata, N. and Hirose, C. and Hofman, D. and Hogan, B. E. and Holland, N. A. and Hollows, I. J. and Holz, D. E. and Honet, L. and {Horton-Bailey}, D. J. and Hough, J. and Hourihane, S. and Howard, N. T. and Howell, E. J. and Hoy, C. G. and Hrishikesh, C. A. and Hsi, P. and Hsieh, H.-F. and Hsieh, H.-Y. and Hsiung, C. and Hsu, S.-H. and Hsu, W.-F. and Hu, Q. and Huang, H. Y. and Huang, Y. and Huang, Y. T. and Huddart, A. D. and Hughey, B. and Hui, V. and Husa, S. and Huxford, R. and Iampieri, L. and Iandolo, G. A. and Ianni, M. and Iannone, G. and Iascau, J. and Ide, K. and Iden, R. and Ierardi, A. and Ikeda, S. and Imafuku, H. and Inoue, Y. and Iorio, G. and Iosif, P. and Iqbal, M. H. and Irwin, J. and Ishikawa, R. and Isi, M. and Isleif, K. S. and Itoh, Y. and Iwaya, M. and Iyer, B. R. and Jacquet, C. and Jacquet, P.-E. and Jacquot, T. and Jadhav, S. J. and Jadhav, S. P. and Jain, M. and Jain, T. and James, A. L. and Jani, K. and Janthalur, N. N. and Jaraba, S. and Jaranowski, P. and Jaume, R. and Javed, W. and Jennings, A. and Jensen, M. and Jia, W. and Jiang, J. and Jin, H.-B. and Johns, G. R. and Johnson, N. A. and {Johnson-McDaniel}, N. K. and Johnston, M. C. and Johnston, R. and Johny, N. and Jones, D. H. and Jones, D. I. and Jones, R. and Jose, H. E. and Joshi, P. and Joshi, S. K. and Joubert, G. and Ju, J. and Ju, L. and Jung, K. and Junker, J. and Juste, V. and Kabagoz, H. B. and Kajita, T. and Kaku, I. and Kalogera, V. and Kalomenopoulos, M. and Kamiizumi, M. and Kanda, N. and Kandhasamy, S. and Kang, G. and Kannachel, N. C. and Kanner, J. B. and KantiMahanty, S. A. and Kapadia, S. J. and Kapasi, D. P. and Karthikeyan, M. and Kasprzack, M. and Kato, H. and Kato, T. and Katsavounidis, E. and Katzman, W. and Kaushik, R. and Kawabe, K. and Kawamoto, R. and Keitel, D. and Kemperman, L. J. and Kennington, J. and Kerkow, F. A. and Kesharwani, R. and Key, J. S. and Khadela, R. and Khadka, S. and Khadkikar, S. S. and Khalili, F. Y. and Khan, F. and Khanam, T. and Khursheed, M. and Khusid, N. M. and Kiendrebeogo, W. and Kijbunchoo, N. and Kim, C. and Kim, J. C. and Kim, K. and Kim, M. H. and Kim, S. and Kim, Y.-M. and Kimball, C. and Kimes, K. and Kinnear, M. and Kissel, J. S. and Klimenko, S. and Knee, A. M. and Knox, E. J. and Knust, N. and Kobayashi, K. and Koehlenbeck, S. M. and Koekoek, G. and Kohri, K. and Kokeyama, K. and Koley, S. and Kolitsidou, P. and Koloniari, A. E. and Komori, K. and Kong, A. K. H. and Kontos, A. and Koponen, L. M. and Korobko, M. and Kou, X. and Koushik, A. and Kouvatsos, N. and Kovalam, M. and Koyama, T. and Kozak, D. B. and Kranzhoff, S. L. and Kringel, V. and Krishnendu, N. V. and Kroker, S. and Kr{\'o}lak, A. and Kruska, K. and Kubisz, J. and Kuehn, G. and Kulkarni, S. and Kulur Ramamohan, A. and Kumar, Achal and Kumar, Anil and Kumar, Praveen and Kumar, Prayush and Kumar, Rahul and Kumar, Rakesh and Kume, J. and Kuns, K. and Kuntimaddi, N. and Kuroyanagi, S. and Kuwahara, S. and Kwak, K. and Kwan, K. and Kwon, S. and Lacaille, G. and Laghi, D. and Laity, A. H. and Lalande, E. and Lalleman, M. and Lalremruati, P. C. and Landry, M. and Lane, B. B. and Lang, R. N. and Lange, J. and Langgin, R. and Lantz, B. and L. Rosa, I. and Larsen, J. and {Lartaux-Vollard}, A. and Lasky, P. D. and Lawrence, J. and Laxen, M. and Lazarte, C. and Lazzarini, A. and Lazzaro, C. and Leaci, P. and Leali, L. and Lecoeuche, Y. K. and Lee, H. M. and Lee, H. W. and Lee, J. and Lee, K. and Lee, R.-K. and Lee, R. and Lee, Sungho and Lee, Sunjae and Lee, Y. and Legred, I. N. and Lehmann, J. and Lehner, L. and Le Jean, M. and Lematre, A. and Lenti, M. and Leonardi, M. and Lequime, M. and Leroy, N. and Lesovsky, M. and Letendre, N. and Lethuillier, M. and Levin, Y. and Leyde, K. and Li, A. K. Y. and Li, K. L. and Li, X. and Li, Y. and Li, Z. and Lihos, A. and Lin, E. T. and Lin, F. and Lin, L. C.-C. and Lin, Y.-C. and Lindsay, C. and Linker, S. D. and Liu, A. and Liu, G. C. and Liu, Jian and Llamas Villarreal, F. and {Llobera-Querol}, J. and Lo, R. K. L. and Locquet, J.-P. and Loggins, S. C. G. and Loizou, M. R. and London, L. T. and Longo, A. and Lopez, D. and Lopez Portilla, M. and Lorenzini, M. and {Lorenzo-Medina}, A. and Loriette, V. and Lormand, M. and Losurdo, G. and Lotti, E. and Lott IV, T. P. and Lough, J. D. and Loughlin, H. A. and Lousto, C. O. and Low, N. and Lu, N. and Lucchesi, L. and L{\"u}ck, H. and Lumaca, D. and Lundgren, A. P. and Lussier, A. W. and Macas, R. and MacInnis, M. and Macleod, D. M. and MacMillan, I. A. O. and Macquet, A. and Maeda, K. and Maenaut, S. and Magare, S. S. and Magee, R. M. and Maggio, E. and Maggiore, R. and Magnozzi, M. and Mahesh, M. and Maini, M. and Majhi, S. and Majorana, E. and Makarem, C. N. and Malakar, D. and {Malaquias-Reis}, J. A. and Mali, U. and Maliakal, S. and Malik, A. and Mallick, L. and Malz, A.-K. and Man, N. and Mancarella, M. and Mandic, V. and Mangano, V. and Mannix, B. and Mansell, G. L. and Manske, M. and Mantovani, M. and Mapelli, M. and Marinelli, C. and Marion, F. and Markosyan, A. S. and Markowitz, A. and Maros, E. and Marsat, S. and Martelli, F. and Martin, I. W. and Martin, R. M. and Martinez, B. B. and Martinez, D. A. and Martinez, M. and Martinez, V. and Martini, A. and Martins, J. C. and Martynov, D. V. and Marx, E. J. and Massaro, L. and Masserot, A. and {Masso-Reid}, M. and Mastrogiovanni, S. and Matcovich, T. and Matiushechkina, M. and Maurin, L. and Mavalvala, N. and Maxwell, N. and McCarrol, G. and McCarthy, R. and McClelland, D. E. and McCormick, S. and McCuller, L. and McEachin, S. and McElhenny, C. and McGhee, G. I. and McGinn, J. and McGowan, K. B. M. and McIver, J. and McLeod, A. and McMahon, I. and McRae, T. and McTeague, R. and Meacher, D. and Meagher, B. N. and Mechum, R. and Meijer, Q. and Melatos, A. and Menoni, C. S. and Mera, F. and Mercer, R. A. and Mereni, L. and Merfeld, K. and Merilh, E. L. and M{\'e}rou, J. R. and Merritt, J. D. and Merzougui, M. and Messick, C. and Mestichelli, B. and {Meyer-Conde}, M. and Meylahn, F. and Mhaske, A. and Miani, A. and Miao, H. and Michel, C. and Michimura, Y. and Middleton, H. and Mihaylov, D. P. and Miller, S. J. and Millhouse, M. and Milotti, E. and Milotti, V. and Minenkov, Y. and Minihan, E. M. and Mir, {\relax Ll}. M. and Mirasola, L. and {Miravet-Ten{\'e}s}, M. and Miritescu, C.-A. and Mishra, A. and Mishra, C. and Mishra, T. and Mitchell, A. L. and Mitchell, J. G. and Mitra, S. and Mitrofanov, V. P. and Mitsuhashi, K. and Mittleman, R. and Miyakawa, O. and Miyoki, S. and Miyoko, A. and Mo, G. and Mobilia, L. and Mohapatra, S. R. P. and Mohite, S. R. and {Molina-Ruiz}, M. and Mondin, M. and Montani, M. and Moore, C. J. and Moraru, D. and More, A. and More, S. and Moreno, C. and Moreno, E. A. and Moreno, G. and Moreso Serra, A. and Morisaki, S. and Moriwaki, Y. and Morras, G. and Moscatello, A. and Mould, M. and Mours, B. and {Mow-Lowry}, C. M. and Muccillo, L. and Muciaccia, F. and Mukherjee, D. and Mukherjee, Samanwaya and Mukherjee, Soma and Mukherjee, Subroto and Mukherjee, Suvodip and Mukund, N. and Mullavey, A. and Mullock, H. and Mundi, J. and Mungioli, C. L. and Murakoshi, M. and Murray, P. G. and Nabari, D. and Nadji, S. L. and Nagar, A. and Nagarajan, N. and Nakagaki, K. and Nakamura, K. and Nakano, H. and Nakano, M. and {Nanadoumgar-Lacroze}, D. and Nandi, D. and Napolano, V. and Narayan, P. and Nardecchia, I. and Narikawa, T. and Narola, H. and Naticchioni, L. and Nayak, R. K. and Negri, L. and Nela, A. and Nelle, C. and Nelson, A. and Nelson, T. J. N. and Nery, M. and Neunzert, A. and Ng, S. and Nguyen Quynh, L. and Nichols, S. A. and Nielsen, A. B. and Nishino, Y. and Nishizawa, A. and Nissanke, S. and Niu, W. and Nocera, F. and Noller, J. and Norman, M. and North, C. and Novak, J. and Nowicki, R. and Nu{\~n}o Siles, J. F. and Nuttall, L. K. and Obayashi, K. and Oberling, J. and O'Dell, J. and Oelker, E. and Oertel, M. and Oganesyan, G. and O'Hanlon, T. and Ohashi, M. and Ohme, F. and Oliveri, R. and Omer, R. and O'Neal, B. and Onishi, M. and Oohara, K. and O'Reilly, B. and Orselli, M. and O'Shaughnessy, R. and O'Shea, S. and Oshino, S. and Osthelder, C. and Ota, I. and Ottaway, D. J. and Ouzriat, A. and Overmier, H. and Owen, B. J. and Ozaki, R. and Pace, A. E. and Pagano, R. and Page, M. A. and Pai, A. and Paiella, L. and Pal, A. and Pal, S. and Palaia, M. A. and P{\'a}lfi, M. and Palma, P. P. and Palomba, C. and Palud, P. and Pan, H. and Pan, J. and Pan, K. C. and Panda, P. K. and Pandey, Shiksha and Pandey, Swadha and Pang, P. T. H. and Pannarale, F. and Pannone, K. A. and Pant, B. C. and Panther, F. H. and Panzeri, M. and Paoletti, F. and Paolone, A. and Papadopoulos, A. and Papalexakis, E. E. and Papalini, L. and Papigkiotis, G. and Paquis, A. and Parisi, A. and Park, B.-J. and Park, J. and Parker, W. and Pascale, G. and Pascucci, D. and Pasqualetti, A. and Passaquieti, R. and Passenger, L. and Passuello, D. and Patane, O. and Patel, A. V. and Pathak, D. and Patra, A. and Patricelli, B. and Patterson, B. G. and Paul, K. and Paul, S. and Payne, E. and Pearce, T. and Pedraza, M. and Pele, A. and Pe{\~n}a Arellano, F. E. and Peng, X. and Peng, Y. and Penn, S. and Penuliar, M. D. and Perego, A. and Pereira, Z. and P{\'e}rigois, C. and Perna, G. and Perreca, A. and Perret, J. and Perri{\`e}s, S. and Perry, J. W. and Pesios, D. and Peters, S. and Petracca, S. and Petrillo, C. and Pfeiffer, H. P. and Pham, H. and Pham, K. A. and Phukon, K. S. and Phurailatpam, H. and Piarulli, M. and Piccari, L. and Piccinni, O. J. and Pichot, M. and Piendibene, M. and Piergiovanni, F. and Pierini, L. and Pierra, G. and Pierro, V. and Pietrzak, M. and Pillas, M. and Pilo, F. and Pinard, L. and Pinto, I. M. and Pinto, M. and Piotrzkowski, B. J. and Pirello, M. and Pitkin, M. D. and Placidi, A. and Placidi, E. and Planas, M. L. and Plastino, W. and Plunkett, C. and Poggiani, R. and Polini, E. and Pomper, J. and Pompili, L. and Poon, J. and Porcelli, E. and Porter, E. K. and Posnansky, C. and Poulton, R. and Powell, J. and Prabhu, G. S. and Pracchia, M. and Pradhan, B. K. and Pradier, T. and Prajapati, A. K. and Prasai, K. and Prasanna, R. and Prasia, P. and Pratten, G. and Principe, G. and Prodi, G. A. and Prosperi, P. and Prosposito, P. and Providence, A. C. and Puecher, A. and Pullin, J. and Puppo, P. and P{\"u}rrer, M. and Qi, H. and Qin, J. and Qu{\'e}m{\'e}ner, G. and Quetschke, V. and Quinonez, P. J. and Qutob, N. and Rading, R. and Rainho, I. and Raja, S. and Rajan, C. and Rajbhandari, B. and Ramirez, K. E. and Ramis Vidal, F. A. and Ramos Arevalo, M. and {Ramos-Buades}, A. and Ranjan, S. and Ransom, K. and Rapagnani, P. and Ratto, B. and Ravichandran, A. and Ray, A. and Raymond, V. and Razzano, M. and Read, J. and Regimbau, T. and Reid, S. and Reissel, C. and Reitze, D. H. and Renzini, A. I. and Revenu, B. and Revilla Pe{\~n}a, A. and Reyes, R. and Ricca, L. and Ricci, F. and Ricci, M. and Ricciardone, A. and Rice, J. and Richardson, J. W. and Richardson, M. L. and Rijal, A. and Riles, K. and Riley, H. K. and Rinaldi, S. and Rittmeyer, J. and Robertson, C. and Robinet, F. and Robinson, M. and Rocchi, A. and Rolland, L. and Rollins, J. G. and Romano, A. E. and Romano, R. and Romero, A. and {Romero-Shaw}, I. M. and Romie, J. H. and Ronchini, S. and Roocke, T. J. and Rosa, L. and Rosauer, T. J. and Rose, C. A. and Rosi{\'n}ska, D. and Ross, M. P. and {Rossello-Sastre}, M. and Rowan, S. and Roy, S. K. and Roy, S. and Rozza, D. and Ruggi, P. and Ruhama, N. and Ruiz Morales, E. and {Ruiz-Rocha}, K. and Sachdev, S. and Sadecki, T. and Saffarieh, P. and {Safi-Harb}, S. and Sah, M. R. and Saha, S. and Sainrat, T. and Sajith Menon, S. and Sakai, K. and Sakai, Y. and Sakellariadou, M. and Sakon, S. and Salafia, O. S. and {Salces-Carcoba}, F. and Salconi, L. and Saleem, M. and Salemi, F. and Sall{\'e}, M. and Salunkhe, S. U. and Salvador, S. and Salvarese, A. and Samajdar, A. and Sanchez, A. and Sanchez, E. J. and Sanchez, L. E. and {Sanchis-Gual}, N. and Sanders, J. R. and S{\"a}nger, E. M. and Santoliquido, F. and Sarandrea, F. and Saravanan, T. R. and Sarin, N. and Sarkar, P. and Sasli, A. and Sassi, P. and Sassolas, B. and Sathyaprakash, B. S. and Sato, R. and Sato, S. and Sato, Yukino and Sato, Yu and Sauter, O. and Savage, R. L. and Sawada, T. and Sawant, H. L. and Sayah, S. and Scacco, V. and Schaetzl, D. and Scheel, M. and Schiebelbein, A. and Schiworski, M. G. and Schmidt, P. and Schmidt, S. and Schnabel, R. and Schneewind, M. and Schofield, R. M. S. and Schouteden, K. and Schulte, B. W. and Schutz, B. F. and Schwartz, E. and Scialpi, M. and Scott, J. and Scott, S. M. and Sedas, R. M. and Seetharamu, T. C. and {Seglar-Arroyo}, M. and Sekiguchi, Y. and Sellers, D. and Sembo, N. and Sengupta, A. S. and Seo, E. G. and Seo, J. W. and Sequino, V. and Serra, M. and Sevrin, A. and Shaffer, T. and Shah, U. S. and Shaikh, M. A. and Shao, L. and Sharma, A. K. and Sharma, Preeti and Sharma, Prianka and Sharma, Ritwik and Sharma Chaudhary, S. and Shawhan, P. and Shcheblanov, N. S. and Sheridan, E. and Shi, Z.-H. and Shikauchi, M. and Shimomura, R. and Shinkai, H. and Shirke, S. and Shoemaker, D. H. and Shoemaker, D. M. and Short, R. W. and ShyamSundar, S. and Sider, A. and Siegel, H. and Sigg, D. and Silenzi, L. and Silvestri, L. and Simmonds, M. and Singer, L. P. and Singh, Amitesh and Singh, Anika and Singh, D. and Singh, N. and Singh, S. and Sintes, A. M. and Sipala, V. and Skliris, V. and Slagmolen, B. J. J. and Slater, D. A. and {Slaven-Blair}, T. J. and Smetana, J. and Smith, J. R. and Smith, L. and Smith, R. J. E. and Smith, W. J. and {Soares de Albuquerque Filho}, S. and {Soares-Santos}, M. and Somiya, K. and Song, I. and Soni, S. and Sordini, V. and Sorrentino, F. and Sotani, H. and Spada, F. and Spagnuolo, V. and Spencer, A. P. and Spinicelli, P. and Srivastava, A. K. and Stachurski, F. and Stark, C. J. and Steer, D. A. and Steinle, N. and Steinlechner, J. and Steinlechner, S. and Stergioulas, N. and Stevens, P. and Stevenson, S. P. and StPierre, M. and Strong, M. D. and Strunk, A. and Stuver, A. L. and Suchenek, M. and Sudhagar, S. and Sudo, Y. and Sueltmann, N. and Suleiman, L. and Sullivan, K. D. and Sun, J. and Sun, L. and Sunil, S. and Suresh, J. and Sutton, B. J. and Sutton, P. J. and Suzuki, K. and Suzuki, M. and Swain, S. and Swinkels, B. L. and Syx, A. and Szczepa{\'n}czyk, M. J. and Szewczyk, P. and Tacca, M. and Tagoshi, H. and Takada, K. and Takahashi, H. and Takahashi, R. and Takamori, A. and Takano, S. and Takeda, H. and Takeshita, K. and Takimoto Schmiegelow, I. and {Takou-Ayaoh}, M. and Talbot, C. and Tamaki, M. and Tamanini, N. and Tanabe, D. and Tanaka, K. and Tanaka, S. J. and Tanioka, S. and Tanner, D. B. and Tanner, W. and Tao, L. and Tapia, R. D. and Tapia San Mart{\'i}n, E. N. and Taranto, C. and Taruya, A. and Tasson, J. D. and Tau, J. G. and Tellez, D. and Tenorio, R. and Themann, H. and Theodoropoulos, A. and Thirugnanasambandam, M. P. and Thomas, L. M. and Thomas, M. and Thomas, P. and Thompson, J. E. and Thondapu, S. R. and Thorne, K. A. and Thrane, E. and Tissino, J. and Tiwari, A. and Tiwari, Pawan and Tiwari, Praveer and Tiwari, S. and Tiwari, V. and Todd, M. R. and Toffano, M. and Toivonen, A. M. and Toland, K. and Tolley, A. E. and Tomaru, T. and Tommasini, V. and Tomura, T. and Tong, H. and {Tong-Yu}, C. and {Torres-Forn{\'e}}, A. and Torrie, C. I. and {Tosta e Melo}, I. and Tournefier, E. and Trad Nery, M. and Tran, K. and Trapananti, A. and Travaglini, R. and Travasso, F. and Traylor, G. and Trevor, M. and Tringali, M. C. and Tripathee, A. and Troian, G. and Trovato, A. and Trozzo, L. and Trudeau, R. J. and Tsang, T. and Tsuchida, S. and Tsukada, L. and Turbang, K. and Turconi, M. and Turski, C. and Ubach, H. and Uchikata, N. and Uchiyama, T. and Udall, R. P. and Uehara, T. and Ueno, K. and Undheim, V. and Uronen, L. E. and Ushiba, T. and Vacatello, M. and Vahlbruch, H. and Vaidya, N. and Vajente, G. and Vajpeyi, A. and Valencia, J. and Valentini, M. and {Vallejo-Pe{\~n}a}, S. A. and Vallero, S. and Valsan, V. and {van Dael}, M. and {Van den Bossche}, E. and {van den Brand}, J. F. J. and Van Den Broeck, C. and {van der Sluys}, M. and {Van de Walle}, A. and {van Dongen}, J. and Vandra, K. and VanDyke, M. and {van Haevermaet}, H. and {van Heijningen}, J. V. and Van Hove, P. and Vanier, J. and VanKeuren, M. and Vanosky, J. and {van Remortel}, N. and Vardaro, M. and Vargas, A. F. and Varma, V. and Vazquez, A. N. and Vecchio, A. and Vedovato, G. and Veitch, J. and Veitch, P. J. and Venikoudis, S. and Venterea, R. C. and Verdier, P. and Vereecken, M. and Verkindt, D. and Verma, B. and Verma, Y. and Vermeulen, S. M. and Vetrano, F. and Veutro, A. and Vicer{\'e}, A. and Vidyant, S. and Viets, A. D. and Vijaykumar, A. and Vilkha, A. and Villanueva Espinosa, N. and {Villa-Ortega}, V. and Vincent, E. T. and Vinet, J.-Y. and Viret, S. and Vitale, S. and Vocca, H. and Voigt, D. and {von Reis}, E. R. G. and {von Wrangel}, J. S. A. and Vossius, W. E. and Vujeva, L. and Vyatchanin, S. P. and Wack, J. and Wade, L. E. and Wade, M. and Wagner, K. J. and Wallace, L. and Wang, E. J. and Wang, H. and Wang, J. Z. and Wang, W. H. and Wang, Y. F. and Waratkar, G. and Warner, J. and Was, M. and Washimi, T. and Washington, N. Y. and Watarai, D. and Weaver, B. and Webster, S. A. and Weickhardt, N. L. and Weinert, M. and Weinstein, A. J. and Weiss, R. and Wen, L. and Wette, K. and Whelan, J. T. and Whiting, B. F. and Whittle, C. and Wickens, E. G. and Wilken, D. and Wilkin, A. T. and Williams, B. M. and Williams, D. and Williams, M. J. and Williams, N. S. and Willis, J. L. and Willke, B. and Wils, M. and Wilson, L. and Winborn, C. W. and Winterflood, J. and Wipf, C. C. and Woan, G. and Woehler, J. and Wolfe, N. E. and Wong, H. T. and Wong, H. W. Y. and Wong, I. C. F. and Wong, K. and Wouters, T. and Wright, J. L. and Wu, B. and Wu, C. and Wu, D. S. and Wu, H. and Wu, K. and Wu, Q. and Wu, Y. and Wu, Z. and Wuchner, E. and Wysocki, D. M. and Xu, V. A. and Xu, Y. and Yadav, N. and Yamamoto, H. and Yamamoto, K. and Yamamoto, T. S. and Yamamoto, T. and Yamazaki, R. and Yan, T. and Yang, K. Z. and Yang, Y. and Yarbrough, Z. and Yebana, J. and Yeh, S.-W. and Yelikar, A. B. and Yin, X. and Yokoyama, J. and Yokozawa, T. and Yuan, S. and Yuzurihara, H. and Zanolin, M. and Zeeshan, M. and Zelenova, T. and Zendri, J.-P. and Zeoli, M. and Zerrad, M. and Zevin, M. and Zhang, L. and Zhang, N. and Zhang, R. and Zhang, T. and Zhao, C. and Zhao, Yue and Zhao, Yuhang and Zhao, Z.-C. and Zheng, Y. and Zhong, H. and Zhou, H. and Zhu, H. O. and Zhu, Z.-H. and Zimmerman, A. B. and Zimmermann, L. and Zucker, M. E. and Zweizig, J.},
  year = 2025,
  month = oct,
  journal = {The Astrophysical Journal Letters},
  volume = {993},
  number = {1},
  pages = {L25},
  publisher = {The American Astronomical Society},
  issn = {2041-8205},
  doi = {10.3847/2041-8213/ae0c9c},
  urldate = {2026-06-08},
  langid = {english}
}

@article{adePlanck2015Results2016,
  title = {Planck 2015 Results - {{XIII}}. {{Cosmological}} Parameters},
  author = {Ade, P. a. R. and Aghanim, N. and Arnaud, M. and Ashdown, M. and Aumont, J. and Baccigalupi, C. and Banday, A. J. and Barreiro, R. B. and Bartlett, J. G. and Bartolo, N. and Battaner, E. and Battye, R. and Benabed, K. and Beno{\^i}t, A. and {Benoit-L{\'e}vy}, A. and Bernard, J.-P. and Bersanelli, M. and Bielewicz, P. and Bock, J. J. and Bonaldi, A. and Bonavera, L. and Bond, J. R. and Borrill, J. and Bouchet, F. R. and Boulanger, F. and Bucher, M. and Burigana, C. and Butler, R. C. and Calabrese, E. and Cardoso, J.-F. and Catalano, A. and Challinor, A. and Chamballu, A. and Chary, R.-R. and Chiang, H. C. and Chluba, J. and Christensen, P. R. and Church, S. and Clements, D. L. and Colombi, S. and Colombo, L. P. L. and Combet, C. and Coulais, A. and Crill, B. P. and Curto, A. and Cuttaia, F. and Danese, L. and Davies, R. D. and Davis, R. J. and de Bernardis, P. and de Rosa, A. and de Zotti, G. and Delabrouille, J. and D{\'e}sert, F.-X. and Valentino, E. Di and Dickinson, C. and Diego, J. M. and Dolag, K. and Dole, H. and Donzelli, S. and Dor{\'e}, O. and Douspis, M. and Ducout, A. and Dunkley, J. and Dupac, X. and Efstathiou, G. and Elsner, F. and En{\ss}lin, T. A. and Eriksen, H. K. and Farhang, M. and Fergusson, J. and Finelli, F. and Forni, O. and Frailis, M. and Fraisse, A. A. and Franceschi, E. and Frejsel, A. and Galeotta, S. and Galli, S. and Ganga, K. and Gauthier, C. and Gerbino, M. and Ghosh, T. and Giard, M. and {Giraud-H{\'e}raud}, Y. and Giusarma, E. and Gjerl{\o}w, E. and {Gonz{\'a}lez-Nuevo}, J. and G{\'o}rski, K. M. and Gratton, S. and Gregorio, A. and Gruppuso, A. and Gudmundsson, J. E. and Hamann, J. and Hansen, F. K. and Hanson, D. and Harrison, D. L. and Helou, G. and {Henrot-Versill{\'e}}, S. and {Hern{\'a}ndez-Monteagudo}, C. and Herranz, D. and Hildebrandt, S. R. and Hivon, E. and Hobson, M. and Holmes, W. A. and Hornstrup, A. and Hovest, W. and Huang, Z. and Huffenberger, K. M. and Hurier, G. and Jaffe, A. H. and Jaffe, T. R. and Jones, W. C. and Juvela, M. and Keih{\"a}nen, E. and Keskitalo, R. and Kisner, T. S. and Kneissl, R. and Knoche, J. and Knox, L. and Kunz, M. and {Kurki-Suonio}, H. and Lagache, G. and L{\"a}hteenm{\"a}ki, A. and Lamarre, J.-M. and Lasenby, A. and Lattanzi, M. and Lawrence, C. R. and Leahy, J. P. and Leonardi, R. and Lesgourgues, J. and Levrier, F. and Lewis, A. and Liguori, M. and Lilje, P. B. and {Linden-V{\o}rnle}, M. and {L{\'o}pez-Caniego}, M. and Lubin, P. M. and {Mac{\'i}as-P{\'e}rez}, J. F. and Maggio, G. and Maino, D. and Mandolesi, N. and Mangilli, A. and Marchini, A. and Maris, M. and Martin, P. G. and Martinelli, M. and {Mart{\'i}nez-Gonz{\'a}lez}, E. and Masi, S. and Matarrese, S. and McGehee, P. and Meinhold, P. R. and Melchiorri, A. and Melin, J.-B. and Mendes, L. and Mennella, A. and Migliaccio, M. and Millea, M. and Mitra, S. and {Miville-Desch{\^e}nes}, M.-A. and Moneti, A. and Montier, L. and Morgante, G. and Mortlock, D. and Moss, A. and Munshi, D. and Murphy, J. A. and Naselsky, P. and Nati, F. and Natoli, P. and Netterfield, C. B. and {N{\o}rgaard-Nielsen}, H. U. and Noviello, F. and Novikov, D. and Novikov, I. and Oxborrow, C. A. and Paci, F. and Pagano, L. and Pajot, F. and Paladini, R. and Paoletti, D. and Partridge, B. and Pasian, F. and Patanchon, G. and Pearson, T. J. and Perdereau, O. and Perotto, L. and Perrotta, F. and Pettorino, V. and Piacentini, F. and Piat, M. and Pierpaoli, E. and Pietrobon, D. and Plaszczynski, S. and Pointecouteau, E. and Polenta, G. and Popa, L. and Pratt, G. W. and Pr{\'e}zeau, G. and Prunet, S. and Puget, J.-L. and Rachen, J. P. and Reach, W. T. and Rebolo, R. and Reinecke, M. and Remazeilles, M. and Renault, C. and Renzi, A. and Ristorcelli, I. and Rocha, G. and Rosset, C. and Rossetti, M. and Roudier, G. and {d'Orfeuil}, B. Rouill{\'e} and {Rowan-Robinson}, M. and {Rubi{\~n}o-Mart{\'i}n}, J. A. and Rusholme, B. and Said, N. and Salvatelli, V. and Salvati, L. and Sandri, M. and Santos, D. and Savelainen, M. and Savini, G. and Scott, D. and Seiffert, M. D. and Serra, P. and Shellard, E. P. S. and Spencer, L. D. and Spinelli, M. and Stolyarov, V. and Stompor, R. and Sudiwala, R. and Sunyaev, R. and Sutton, D. and {Suur-Uski}, A.-S. and Sygnet, J.-F. and Tauber, J. A. and Terenzi, L. and Toffolatti, L. and Tomasi, M. and Tristram, M. and Trombetti, T. and Tucci, M. and Tuovinen, J. and T{\"u}rler, M. and Umana, G. and Valenziano, L. and Valiviita, J. and Tent, F. Van and Vielva, P. and Villa, F. and Wade, L. A. and Wandelt, B. D. and Wehus, I. K. and White, M. and White, S. D. M. and Wilkinson, A. and Yvon, D. and Zacchei, A. and Zonca, A.},
  year = 2016,
  month = oct,
  journal = {Astronomy \& Astrophysics},
  volume = {594},
  pages = {A13},
  publisher = {EDP Sciences},
  issn = {0004-6361, 1432-0746},
  doi = {10.1051/0004-6361/201525830},
  urldate = {2026-05-07},
  copyright = {\copyright{} ESO, 2016},
  langid = {english}
}

@article{aiLargeSkyArea2016,
  title = {The {{Large Sky Area Multi-object Fiber Spectroscopic Telescope Quasar Survey}}: {{Quasar Properties}} from the {{First Data Release}}},
  shorttitle = {The {{Large Sky Area Multi-object Fiber Spectroscopic Telescope Quasar Survey}}},
  author = {Ai, Y. L. and Wu, Xue-Bing and Yang, Jinyi and Yang, Qian and Wang, Feige and Guo, Rui and Zuo, Wenwen and Dong, Xiaoyi and Zhang, Y. -X. and Yuan, H. -L. and Song, Y. -H. and Wang, Jianguo and Dong, Xiaobo and Yang, M. and {-Wu}, H. and Shen, S. -Y. and Shi, J. -R. and He, B. -L. and Lei, Y. -J. and Li, Y. -B. and Luo, A. -L. and Zhao, Y. -H. and Zhang, H. -T.},
  year = 2016,
  month = feb,
  journal = {The Astronomical Journal},
  volume = {151},
  pages = {24},
  publisher = {IOP},
  issn = {0004-6256},
  doi = {10.3847/0004-6256/151/2/24},
  urldate = {2025-06-21},
  langid = {american}
}

@article{antoniniMERGINGBLACKHOLE2016,
  title = {{{MERGING BLACK HOLE BINARIES IN GALACTIC NUCLEI}}: {{IMPLICATIONS FOR ADVANCED-LIGO DETECTIONS}}},
  shorttitle = {{{MERGING BLACK HOLE BINARIES IN GALACTIC NUCLEI}}},
  author = {Antonini, Fabio and Rasio, Frederic A.},
  year = 2016,
  month = nov,
  journal = {The Astrophysical Journal},
  volume = {831},
  number = {2},
  pages = {187},
  publisher = {The American Astronomical Society},
  issn = {0004-637X},
  doi = {10.3847/0004-637X/831/2/187},
  urldate = {2026-05-21},
  langid = {english}
}

@article{ashtonBilbyUserfriendlyBayesian2019,
  title = {Bilby: {{A User-friendly Bayesian Inference Library}} for {{Gravitational-wave Astronomy}}},
  shorttitle = {Bilby},
  author = {Ashton, Gregory and H{\"u}bner, Moritz and Lasky, Paul D. and Talbot, Colm and Ackley, Kendall and Biscoveanu, Sylvia and Chu, Qi and Divakarla, Atul and Easter, Paul J. and Goncharov, Boris and Vivanco, Francisco Hernandez and Harms, Jan and Lower, Marcus E. and Meadors, Grant D. and Melchor, Denyz and Payne, Ethan and Pitkin, Matthew D. and Powell, Jade and Sarin, Nikhil and Smith, Rory J. E. and Thrane, Eric},
  year = 2019,
  month = apr,
  journal = {The Astrophysical Journal Supplement Series},
  volume = {241},
  number = {2},
  pages = {27},
  publisher = {The American Astronomical Society},
  issn = {0067-0049},
  doi = {10.3847/1538-4365/ab06fc},
  urldate = {2026-05-08},
  langid = {english}
}

@article{ashtonCurrentObservationsAre2021,
  title = {Current Observations Are Insufficient to Confidently Associate the Binary Black Hole Merger {{GW190521}} with {{AGN J124942}}.3+344929},
  author = {Ashton, Gregory and Ackley, Kendall and Hernandez, Ignacio Maga{\~n}a and Piotrzkowski, Brandon},
  year = 2021,
  month = dec,
  journal = {Classical and Quantum Gravity},
  volume = {38},
  number = {23},
  eprint = {2009.12346},
  primaryclass = {astro-ph},
  pages = {235004},
  issn = {0264-9381, 1361-6382},
  doi = {10.1088/1361-6382/ac33bb},
  urldate = {2024-05-11},
  archiveprefix = {arXiv},
  langid = {american}
}

@article{astropycollaborationAstropyCommunityPython2013,
  title = {Astropy: {{A}} Community {{Python}} Package for Astronomy},
  shorttitle = {Astropy},
  author = {{Astropy Collaboration} and Robitaille, Thomas P. and Tollerud, Erik J. and Greenfield, Perry and Droettboom, Michael and Bray, Erik and Aldcroft, Tom and Davis, Matt and Ginsburg, Adam and {Price-Whelan}, Adrian M. and Kerzendorf, Wolfgang E. and Conley, Alexander and Crighton, Neil and Barbary, Kyle and Muna, Demitri and Ferguson, Henry and Grollier, Fr{\'e}d{\'e}ric and Parikh, Madhura M. and Nair, Prasanth H. and Unther, Hans M. and Deil, Christoph and Woillez, Julien and Conseil, Simon and Kramer, Roban and Turner, James E. H. and Singer, Leo and Fox, Ryan and Weaver, Benjamin A. and Zabalza, Victor and Edwards, Zachary I. and Azalee Bostroem, K. and Burke, D. J. and Casey, Andrew R. and Crawford, Steven M. and Dencheva, Nadia and Ely, Justin and Jenness, Tim and Labrie, Kathleen and Lim, Pey Lian and Pierfederici, Francesco and Pontzen, Andrew and Ptak, Andy and Refsdal, Brian and Servillat, Mathieu and Streicher, Ole},
  year = 2013,
  month = oct,
  journal = {Astronomy and Astrophysics},
  volume = {558},
  pages = {A33},
  issn = {0004-6361},
  doi = {10.1051/0004-6361/201322068},
  urldate = {2025-04-18}
}

@article{bartosRapidBrightStellarmass2017,
  title = {Rapid and {{Bright Stellar-mass Binary Black Hole Mergers}} in {{Active Galactic Nuclei}}},
  author = {Bartos, Imre and Kocsis, Bence and Haiman, Zolt{\'a}n and M{\'a}rka, Szabolcs},
  year = 2017,
  month = jan,
  journal = {The Astrophysical Journal},
  volume = {835},
  number = {2},
  pages = {165},
  publisher = {The American Astronomical Society},
  issn = {0004-637X},
  doi = {10.3847/1538-4357/835/2/165},
  urldate = {2025-01-17},
  langid = {english}
}

@article{baveraImpactMasstransferPhysics2021,
  title = {The Impact of Mass-Transfer Physics on the Observable Properties of Field Binary Black Hole Populations},
  author = {Bavera, Simone S. and Fragos, Tassos and Zevin, Michael and Berry, Christopher P. L. and Marchant, Pablo and Andrews, Jeff J. and Coughlin, Scott and Dotter, Aaron and Kovlakas, Konstantinos and Misra, Devina and {Serra-Perez}, Juan G. and Qin, Ying and Rocha, Kyle A. and {Rom{\'a}n-Garza}, Jaime and Tran, Nam H. and Zapartas, Emmanouil},
  year = 2021,
  month = mar,
  journal = {Astronomy \& Astrophysics},
  volume = {647},
  pages = {A153},
  issn = {0004-6361, 1432-0746},
  doi = {10.1051/0004-6361/202039804},
  urldate = {2026-06-14},
  copyright = {https://www.edpsciences.org/en/authors/copyright-and-licensing}
}

@article{bellovaryMIGRATIONTRAPSDISKS2016,
  title = {{{MIGRATION TRAPS IN DISKS AROUND SUPERMASSIVE BLACK HOLES}}},
  author = {Bellovary, Jillian M. and Low, Mordecai-Mark Mac and McKernan, Barry and Ford, K. E. Saavik},
  year = 2016,
  month = mar,
  journal = {The Astrophysical Journal Letters},
  volume = {819},
  number = {2},
  pages = {L17},
  publisher = {The American Astronomical Society},
  issn = {2041-8205},
  doi = {10.3847/2041-8205/819/2/L17},
  urldate = {2026-06-15},
  langid = {english}
}
\bibliographystyle{aasjournalv7}
\end{document}